\documentclass[hyper]{JHEP3}
\newcommand{\Scal}[1]{\Bigl ({#1} \Bigr )}
\newcommand{\scal}[1]{\bigl ({#1} \bigr )}
\def\ie{{\it i.e.}\ }
\def\eg{{\it e.g.}\ }

\DeclareMathAlphabet{\mathpzc}{OT1}{pzc}{m}{it}
\usepackage{dsfont}
\usepackage[Symbol]{upgreek}
\usepackage{amssymb}
\usepackage{amsmath}

\newcommand{\ord}[1]{{\scriptscriptstyle (#1)}}

\def\cN{{\mathcal{N}}}
\def\gl{\mathfrak{gl}}
\def\sl{\mathfrak{sl}}
\def\so{\mathfrak{so}}
\def\su{\mathfrak{su}}

\def\e{\mathfrak{e}}

\def\bea{\begin{eqnarray}}
\def\eea{\end{eqnarray}}
\def\be{\begin{equation}}
\def\ee{\end{equation}}

\newcommand{\CR}{\nonumber \\*}
\def\nn{\nonumber}
\def\cH{{\mathcal H}}
\def\cV{{\mathcal V}}
\def\cK{{\mathcal{K}}}

\newcommand{\pA}{{\text{\tiny A}}}

\def\cE{{\mathcal{E}}}
\def\un{{\mathpzc{1}}}
\def\deux{{\mathpzc{2}}}
\def\trois{{\mathpzc{3}}}

\title{A bubbling bolt}

\author{Guillaume Bossard$^{\sf{a}}$  and Stefanos Katmadas$^{\sf{b}}$
\\ {$\sf{a}$ Centre de Physique Th\'eorique, \'Ecole Polytechnique, CNRS, 91128
Palaiseau, France}
\\ {$\sf{b}$ Dipartimento di Fisica, Universit\'a di Milano-Bicocca, I-20126 Milano, Italy }
\\ \email{guillaume.bossard [at] cpht.polytechnique.fr},\\
   \email{stefanos.katmadas  [at] unimib.it}}

\abstract{
We present a new solvable system, solving the equations of five-dimensional ungauged $\cN\!=\!1$ supergravity
coupled to vector multiplets, that allows for non-extremal solutions and reduces to a known system when restricted to the floating brane Ansatz. A two-centre globally hyperbolic smooth geometry is obtained as a solution to this system, describing a bubble linking  a Gibbons--Hawking centre to a charged bolt. However this solution turns out to violate the BPS bound, and we show that its generalisation to an arbitrary number of Gibbons--Hawking centres never admits a spin structure.
}
\preprint{ {CPHT-RR034.0514} }


\begin{document}

\section{Introduction}

One of the most interesting recent developments in black hole physics in the context of string theory
has been the fuzzball proposal of Mathur and collaborators \cite{Mathur:2005ai,Mathur:2005zp}. The main claim is that 
quantum effects modify the dynamics in the background of a black hole at the scale of its horizon, even if the Riemann 
tensor components are all very small compared to the Planck scale in this region. This is rather intuitive for extremal
black holes, whose event horizon is in causal contact with the curvature singularity, whereas non-extremal black holes 
generically admit both an inner and an outer horizon. The quantum state defining the black hole is a superposition of 
pure states, which are in one to one correspondence with pure microstates of the dual conformal field theory. In its 
simpler form, the proposal asserts that, at least in some regime, these microstates have a well defined semi-classical 
limit at strong coupling, and correspond to smooth globally hyperbolic geometries in supergravity. The black hole state,
defined as a distribution over the space of metrics, is then well approximated by a sum of Dirac distributions picked 
at 
stationary metric configurations that describe smooth horizon-less geometries with the same asymptotic structure as 
the corresponding black hole geometry. In a first approximation, the quantum dynamics is described by an uncorrelated
superposition of unitary evolutions in the background of these globally hyperbolic smooth geometries, and the
information paradox is resolved. In this picture, the black hole horizon is replaced by an effective distance at which
the various microstates start to diverge from the classical black hole geometry. If all microstates had a well defined
semi-classical limit as smooth geometries, counting them in the appropriate way would ultimately count the number of
conformal field theory microstates associated to the black hole, and match the exponential of the black hole entropy.

This has resulted in a considerable amount of work and a variety of explicit microstate geometries,
corresponding to BPS 
\cite{Mathur:2003hj,Bena:2005va,Berglund:2005vb,Saxena:2005uk,Bena:2007kg,Bena:2010gg,Lunin:2012gp} 
and non-BPS extremal black holes \cite{Bobev:2011kk,Niehoff:2013mla}. However, extremal black holes do not carry a 
temperature by definition, so that there is no actual information paradox in this case. It is therefore very important
to test the fuzzball proposal in the case of non-extremal black holes. There are only very few non-extremal 
solutions \cite{Jejjala:2005yu,Bena:2009qv,Bobev:2009kn,Giusto:2007tt,Compere:2009iy,Banerjee:2014hza}, what is 
essentially due to the fact that they are all single-centre. In particular, the running bolt solutions 
\cite{Bena:2009qv,Bobev:2009kn} were obtained within a particular solvable system derived from the so-called floating 
brane Ansatz \cite{Bena:2009fi} for ungauged supergravity in five dimensions. This system is a non-extremal deformation of
the known extremal systems and is based on solutions of the Euclidean Maxwell--Einstein equations. For brevity, we
shall refer to it as the floating brane system in this paper. In the special case when the Euclidean base space is
Gibbons--Hawking, the floating brane system reduces either to the BPS system \cite{Behrndt:1997ny} or to the almost-BPS
system \cite{Goldstein:2008fq,Bena:2009ev,Bena:2009en}, which describes extremal non-BPS solutions. Given that
multi-centre BPS and almost-BPS solutions are well studied, it is natural to consider the possibility of constructing
non-extremal multi-centre solutions by generalising known results from the extremal case.

In this paper we consider exactly this possibility and present an explicit two-centre non-extremal
solution to five-dimensional $\cN\!=\!1$ ungauged supergravity coupled to vector multiplets that is
in particular a solution to the floating brane system of \cite{Bena:2009fi}. As one expects based on the intuition that
non-extremal black holes always attract and therefore cannot form stable bound states, the solution
we present does not feature a horizon and is in fact smooth and free of closed time-like curves. This solution includes 
both a running bolt homology sphere and a Gibbons--Hawking centre, which together define a bubble homology sphere 
supporting fluxes. The system generalises to an arbitrary number of Gibbons--Hawking centres, and one therefore expects 
to be able to obtain a large class of smooth solutions in this way.

Although these two-centre solutions have naively the asymptotic charges of a non-extremal four-dimensional black hole, similar to
\cite{Bena:2009qv,Bobev:2009kn,Compere:2009iy}, one finds that the asymptotic charges are beyond the range in which a
regular black hole solution exists, and in particular that the ADM mass of the solution violates the BPS bound, as in
\cite{Compere:2009iy}. As explained in \cite{Gibbons:2013tqa}, this somehow unexpected property arises because of the
failure of these smooth space-times to admit a spin structure. We shall indeed show that a homology sphere linking a
bolt cycle and a Gibbons--Hawking centre supports a non-vanishing second Stiefel--Whitney class. 

Our strategy is based on the three-dimensional non-linear sigma model obtained by time-like dimensional reduction
\cite{Breitenlohner:1987dg}, and consists in modifying the construction of the almost-BPS system \cite{Bossard:2013oga}
to obtain a non-extremal system that is shown to reproduce the one derived from the floating brane Ansatz in a
particular case. While this system is shown to contain all known extremal systems in various limits, we show that its
non-extremal solutions lie in a different duality orbit from the orbit of regular black hole solutions. This implies that the system does not include solutions
with the same asymptotic charges and asymptotic momenta of the scalars as a regular non-extremal four-dimensional black hole. Therefore, even for the
solutions admitting a spin structure (as $\eg$ the running bolt based on a Kerr--Newman Euclidean electro-vacuum
\cite{Bena:2009qv,Bobev:2009kn}), and preserving the extremality bound, the flow of the scalar fields at infinity is always off the attractor flow associated to a regular black hole. We conclude that any globally hyperbolic smooth solution in these systems, necessarily deviates from regular black hole solutions with the same asymptotic charges all the way to infinity. It implies that if one can interpret them as black hole microstate geometries, any observer anywhere in space-time will be subjected to some kind of fuzziness.

This paper is organised as follows. In section \ref{sec:EinMax-top} we give a self-contained discussion of the
axisymmetric Euclidean Maxwell--Einstein equations, using the formulation in terms of the Ernst potentials. We
extend the known list of instanton solutions to this system by considering multi-centre solutions where exactly one
centre is the Euclidean continuation of a Reissner--Nordstr\"{o}m black hole with a NUT charge, \ie a bolt, while the
remaining centres are taken up by standard Gibbons--Hawking nuts. An extended discussion on the regularity of these
solutions is given, where we also show that they do not admit a spin structure. In section \ref{sec:NE-sys} we
proceed to define the non-extremal system in terms of the non-linear sigma model in three dimensions and analyse its
properties, including the various extremal limits, the five-dimensional uplift and the asymptotic
structure. Section \ref{sec:2cen-ex} is devoted to the presentation of the explicit two-centre globally hyperbolic
smooth geometry and its properties, including a description of the bubble structure that parallels the one in
\cite{Bena:2005va,Berglund:2005vb}. We verify explicitly the violation of the BPS bound in a large class of solutions. 
Finally, we conclude in section \ref{sec:concl} with a discussion of further directions.

\section{Maxwell--Einstein instantons}
\label{sec:EinMax-top}

In this paper we discuss solutions of four-dimensional $\cN=2$ supergravity coupled to $n_v$ vector multiplets, which
can be obtained in partially solvable systems. All these solutions lift by construction to $\cN=1$ supergravity in five 
dimensions. These partially solvable systems include a solvable system in the background of a solution to Euclidean 
Maxwell--Einstein equations and we therefore devote this section to a discussion of solutions to these equations.
Depending of the specific uplift we choose to five dimensions, the Euclidean metric solving Maxwell--Einstein equations
is not necessarily realised geometrically in five-dimensions. However this metric will describe the Euclidean base
metric of the five-dimensional space-time in the specific solutions we discuss in this paper. Therefore it will be
important for us to discuss in addition the regularity of these Riemannian manifolds. In section \ref{sec:EinMax} we
give the equations of motion and basic properties of the Euclidean Maxwell--Einstein system, in terms of the
split-complex Ernst potentials. Starting with the general static non-extremal single-centre solution, we then consider
the addition of an arbitrary number of extremal centres in section \ref{sec:instanton}. Finally, we analyse the
regularity conditions for the resulting solutions, demanding that they be everywhere smooth.

\subsection{Euclidean Maxwell--Einstein equations}
\label{sec:EinMax}

Stationary solutions to the Euclidean Maxwell--Einstein equations of motion can be conveniently
recast in terms of scalar variables, known as Ernst potentials. Consider a
metric with an isometry along the direction of the coordinate $\psi$, as
\begin{equation}
ds_4^{\; 2} = V^{-1}( d \psi + w^0 )^2 + V\, \gamma_{ij} dx^i dx^j \,,
\end{equation} 
as well as a gauge field strength invariant under the same isometry, as
\begin{equation}\label{eq:inst-gaug-gen}
F = d( K_+ + K_-)\wedge( d \psi + w^0 ) + V\,\star d(K_+ - K_-)\,.
\end{equation}
In the relations above, $\gamma_{ij}$ is a metric describing a three-dimensional base space and $\star$ is the Hodge 
dual with respect to that metric. Finally,
$V$, $K_\pm$ are scalar functions, while $w^0$ is a vector field, all defined
over the three-dimensional base. The above background solves both the Maxwell
equations and the components of the Einstein equations along $d\psi$ if one defines the functions
$\cE_\pm$ as
\begin{align}
V^{-1}=&\, \cE_+ + \cE_- + K_+K_-\,, \qquad
\star dw^0 = V^2\, (d \cE_+ - d \cE_- + K_- d K_+ - K_+ d K_-)\,, \label{DefVw} 
\end{align}
and imposes that $\cE_\pm$, $K_\pm$ are solutions of
\begin{eqnarray} \label{eq:EK-eoms}
\scal{ \cE_+ + \cE_- + K_+ K_- } \Delta \cE_\pm   &=& 2 ( \nabla \cE_\pm +  K_\mp \nabla K_\pm ) \nabla \cE_\pm \CR
 \scal{ \cE_+ + \cE_- + K_+ K_- } \Delta K_\pm   &=& 2 ( \nabla \cE_\pm +  K_\mp \nabla K_\pm ) \nabla K_\pm \ . 
\end{eqnarray}
The only remaining equation is the three-dimensional base Einstein equation, which reads
\begin{equation}\label{eq:R-base}
R(\gamma)_{ij} =  \frac{ ( \partial_{(i} \cE_+ + K_- \partial_{(i} K_+ ) (  \partial_{j)} \cE_- + K_+  \partial_{j)}  
K_- )}{( \cE_+  + \cE_- + K_+ K_- )^2} - \frac{\partial_{(i} K_+   \partial_{j)} K_- }{ \cE_+  + \cE_- + K_+ K_- }    \ 
,
\end{equation}
and specifies the metric $\gamma_{ij}$. 

It is important to stress that the four scalar fields $\cE_\pm,\, K_{\pm}$ can be shown to parametrise the coset
$SL(3,\mathds{R})/GL(2,\mathds{R})$, so that the above equations are invariant with respect to $SL(3,\mathds{R})$
Harrison transformations. Because the coset component splits in two irreducible representations of $GL(2,\mathds{R})$,
the Harrison transformations only mix $\cE_+$ with $K_+$ and respectively $\cE_-$ with $K_-$. 

There are various interesting extremal limits of the above system, obtained by setting the Ricci tensor to vanish,
so that the three-dimensional base space is flat, \ie  $\gamma_{ij}=\delta_{ij}$. There are five classes of such
solutions, defining five distinct orbits of $SL(3,\mathds{R})$, for which only two out of the four potentials are
non-constant, and are then determined in terms of two arbitrary harmonic functions. Starting from trivial $K_\pm$,
one may set either of the $\cE_\pm$ to a constant, implying that the full four-dimensional curvature is selfdual
for $\cE_-\!=\,$const. or anti-selfdual for $\cE_+\!=\,$const. and that the metric belongs to the class of selfdual
instantons of \cite{Gibbons:1979zt}. For each of these two choices, one may further add a selfdual or anti-selfdual
flux, implying that only one of the $K_\pm$ may be nontrivial. These account for four of the possibilities with a
flat three-dimensional base, the fifth one corresponding to the Israel--Wilson class of solutions, where one sets
both $\cE_\pm=0$ and both $(K_\pm)^{-1}$ are harmonic functions. 

In this paper we are interested in non-extremal solutions, which may not have a flat three-dimensional base. This
is the case with the general Reissner--Nordstr\"{o}m--NUT instanton, described by the functions
\begin{align}\label{eq:RN-nut}
 \cE_\pm = &\, \frac{r - m_{\pm}}{r + m_{\pm}}\,,
 \qquad
 K_{\pm} =  \frac{2 \,e_\pm}{r + m_{\pm}}\,,
\end{align}
where $m_\pm$, $e_\pm$ are constants. The Euclidean four-dimensional metric is
\begin{equation}\label{eq:base-metr}
ds_4^{\; 2} =\,V^{-1}\, ( d\psi + w^0 )^2 
+ V\, \left( dr^2 + ( r^2 -c^2 ) \scal{ d\theta^2 + \sin^2\theta d\varphi^2 } \right)\,,
\end{equation}
where the scale factor V and the one-form $w^0$ are defined according to \eqref{DefVw}, \ie 
\begin{align}
V=&\, \frac{(r+m_+)(r+m_-)}{2\,(r^2-c^2)}\,, \qquad
 w^0 = \frac{m_+-m_-}2\, \scal{\pm 1-\cos \theta} d\varphi \,,
\end{align}
while the non-extremality parameter, $c$, is given by
\begin{equation}\label{eq:c-def}
c^2 = m_+ m_{-} - 2 e_+ e_-\,.
\end{equation}
One can consider more general single-centre examples, obtained for example by the Euclidean continuation
of Kerr solutions. However, the analytic continuation implies that the angular momentum
parameter is pure imaginary, so that the three-dimensional base metric of the associated instanton is not the
same as the one of the Minkowski signature Kerr solution. However we do not consider such examples in this paper.

\subsection{Multi-centre non-extremal base}\label{sec:instanton}

The solutions of the Euclidean Maxwell--Einstein equations are not limited to the solutions one can obtain by analytic 
continuation of solutions defined in Minkowski signature. This observation allows to bypass the standard uniqueness
theorem and define regular multi-centre solutions, including a non-extremal bolt and arbitrary many Gibbons--Hawking
like centres.

Motivated by studies in the probe approximation \cite{Bena:2011fc,Bena:2012zi}, indicating that it should be possible
to add a supersymmetric Gibbons--Hawking centre in the background of a non-extremal geometry, we consider the
generalisation of the BPS system with $\cE_-$ and $K_-$ kept constant to the case where they are the ones of a
single-centre non-extremal static solution \eqref{eq:RN-nut} as in the last section, \ie 
\begin{align}
 \cE_- = &\, \frac{r - m_{-}}{r + m_{-}}\,,
 \qquad
 K_{-} =  \frac{2 \,e_-}{r + m_{-}}\, . \label{NonExtrMinus}
\end{align}
If $\cE_-$ and $K_-$ were constant, $\cE_+$ and $K_+$ would be determined in terms of two arbitrary harmonic functions 
$V$ and $V K_+$. Working out the equations of motion for $\cE_+$ and $K_+$ in this background \eqref{NonExtrMinus}, one 
finds that the two equations are compatible if and only if   
\begin{equation}\cE_+ = -1 + \frac{2}{m_{-}+\frac{c^2}{r}} \scal{ m_{-} -  e_- K_+ } \ . \end{equation}
In this case one gets a solution to the system, provided $K_+$ satisfies
\begin{equation}\Delta K_+ - 2 c^2 \frac{r + m_{-} }{(r^2-c^2) \scal{m_{-} r + c^2 }} \nabla r \cdot \nabla K_+ = - 2 
\frac{e_-}{m_{-}-e_- K_+} \nabla K_+ \cdot \nabla K_+ \ . \label{Phim} \end{equation}
Of course this is solved by \eqref{eq:RN-nut}-\eqref{eq:c-def}, but this is not the
only solution if one does not require spherical symmetry. At large $r$ the equation simplifies,
and one finds that $K_+$ is a solution for any harmonic function ${\cal H}$ such that
\begin{align}
K_+ = \frac{m_-}{e_-} \Scal{1 - \frac{1}{{\cal H}}} \ ,
\qquad
\cE_+  = -1 + 2 \frac{r}{r+ \frac{c^2}{m_-}} \frac{1}{\cH} \ ,
\end{align}
while the scale factor takes the form
\begin{equation}\cE_+ + \cE_- + K_+ K_- = \frac{ r^2-c^2}{(r+m_-)(r + \frac{c^2}{m_-})} \frac{2}{\cH} \ . \end{equation}
These relations can be extended to the full geometry, using them as an Ansatz in \eqref{Phim}, to obtain
a linear equation for ${\cal H}$, as
\begin{equation} \label{eq:Lapl-H}
\Delta {\cal H} =  2 \nabla \Scal{ \mbox{ln}\frac{\sqrt{ r^2 - c^2}}{  m_- r + c^2 }} \cdot \nabla {\cal H} \ . 
\end{equation}
With this Ansatz, one can check that the three-dimensional energy-momentum tensor does not depend on $\cH$ and is
such that
\begin{equation} \frac{ ( \partial_{(i} \cE_+ + K_- \partial_{(i} K_+ ) (  \partial_{j)} \cE_- + K_+  \partial_{j)}  K_- )}{( \cE_+  
+ \cE_- + K_+ K_- )^2} - \frac{\partial_{(i} K_+   \partial_{j)} K_- }{ \cE_+  + \cE_- + K_+ K_- } =  \frac{c^2}{(r^2 - 
c^2)^2 } \partial_{(i}r \partial_{j)} r \ , \end{equation}
and so the three-dimensional base metric is exactly the same as in \eqref{eq:base-metr} above.

In order to solve \eqref{eq:Lapl-H}, it is convenient to use Weyl coordinates on the three-dimensional base,
so that \eqref{eq:base-metr} can be rewritten as 
\begin{eqnarray}  \label{3Dbase} 
\gamma_{ij} dx^i dx^j &=& dr^2 + ( r^2 -c^2 ) \scal{ d\theta^2 + \sin^2\theta d\varphi^2 }  
\CR
 &=& \frac{ r^2 - c^2 }{r_+ r_- }( dz^2 + d\rho^2) + \rho^2 d\varphi^2 \ ,
\end{eqnarray}
where
\begin{equation}r_\pm = \sqrt{ \rho^2 + (z\pm c)^2 } \ , \qquad 2 r = r_+ + r_- \  , \qquad 2 c \cos\theta = r_+ - r_- \ . \end{equation}
In these coordinates, it is straightforward to verify that the function
\begin{eqnarray} \label{eq:H1}
\cH_\pA &=& 
\frac{2\,k\,n_\pA}{r_\pA} \frac{\sqrt{ (R_\pA^{\; 2}  - c^2)\,(r^2 - c^2) + c^2\, r_\pA^{\; 2} }}{(|R_\pA|+m_-)\,(r + 
\frac{c^2}{m_-})} \CR
&=& \frac{2\,k\,n_\pA}{(|R_\pA|+m_-)\,(r + 
\frac{c^2}{m_-})} \frac{ |R_\pA r - c^2 \cos \theta|}{\sqrt{ (R_\pA-r\cos\theta)^2 + (r^2 - c^2) \sin^2\theta}}  \ , 
\end{eqnarray}
solves \eqref{eq:Lapl-H}, where $r_\pA$ is the distance from a second centre
\begin{equation}r_\pA = \sqrt{  \rho^2 + (z-R_\pA)^2 } \ , \label{DefRA}  \end{equation}
at a distance $|R_\pA|>c$ from the origin in cylindrical coordinates. Here, $k\,n_\pA$ stands for the integration
constant, which we have written in this specific form for reasons that will become clear when we discuss regularity,
which implies that $n_\pA$ must necessarily be an integer, whereas the constant $k$ defines a scale that determines
the periodicity of the $\psi$ coordinate. Near $r_\pA=0$, the solution reduces to a Gibbons--Hawking instanton of integral Kaluza--Klein monopole charge
 $n_\pA$, given that $\cH_\pA$ has a single pole at $r_\pA=0$
and that $\cE_-$ and $K_-$ are regular at this point.

Using the linearity of the differential equation \eqref{eq:Lapl-H}, one can consider
the straightforward multi-centre generalisation to
\begin{equation}\label{eq:HA}
\cH =  \frac{r +m_+}{  r + \frac{c^2}{m_-} } 
- \sum_\pA \frac{2\,k\,n_\pA}{r_\pA} 
  \frac{\sqrt{ (R_\pA^{\; 2}  - c^2)\,(r^2 - c^2) + c^2\, r_\pA^{\; 2} }}{(|R_\pA|+m_-)\,(r + \frac{c^2}{m_-})} \ , 
\end{equation}
which describes a Gibbons--Hawking multi-instanton in the background of a single non-extremal centre. In the limit 
$c=m_-=0$, the solution reduces to a Gibbons--Hawking multi-instanton \cite{Gibbons:1979zt}. Note that although the 
expression \eqref{DefRA} is only valid for an axisymmetric solution, the generalisation to arbitrary non-axisymmetric 
solutions with Gibbons--Hawking centres located at $\vec{x}=\vec{x}_\pA$ can straightforwardly be obtained by
viewing $(r,\,\theta,\,\phi)$ as standard polar coordinates and changing to Cartesian coordinates $x^i$, so that
the metric becomes 
\begin{equation}\gamma_{ij} = \Scal{ 1-\frac{c^2}{|x|^2}} \delta_{ij} + \frac{c^2}{|x|^4} x_i x_j \ . \end{equation}
In these coordinates, we have 
\begin{equation}r = |x| \ , \qquad r_\pA = \sqrt{ |x-x_\pA|^2 + c^2  \frac{(\vec{x}_\pA\hspace{-0.5mm} \cdot \vec{x})^2-|x|^2 |x_\pA|^2}{|x|^2 |x_\pA|^2}} \ 
, \end{equation}
for which the bolt is located on the sphere $|x|^2 = c^2$, such that $|x|\ge c$.
Using the definitions above, we find the following functions specifying the Maxwell--Einstein base
\begin{gather}\label{eq:SL3-fun-mult}
 \cE_+ = -1 -\frac{ 2\, m_{-} r }{( c^2+m_{-} r) \sum_\pA \cH_\pA - m_{-} (m_{+}+r)} \,,
 \qquad
 \cE_{-}=\frac{r - m_{-}}{r + m_{-}}
\nn\\
 K_+ = \frac{m_-}{e_-}\,\left(1+\frac{ c^2+m_{-} r }{( c^2+m_{-} r) \sum_\pA \cH_\pA - m_{-} (m_{+}+r)}\right)\,,
 \qquad
 K_{-}=\frac{2\, e_{-}}{r + m_{-}}\,.
\end{gather}
The metric is defined as in  \eqref{eq:base-metr}, with the scaling factor
\begin{equation}
\label{eq:V-mult}
 V = \frac{(m_- +r) \left(m_{-} (m_{+}+r)-( c^2+m_{-} r) \sum_\pA \cH_\pA  \right)}{2 m_- \scal{r^2 -c^2}} \end{equation}
and the Kaluza--Klein vector
\begin{align}\label{eq:w0gen}
w^0 =&\,-\frac{1}{2} \biggl( (m_+-m_-) \cos\theta 
- \sum_\pA\frac{\scal{1-\frac{c^2}{m_-^{\; 2}}}( c^2 + m_- r ) ( r - R_\pA \cos\theta)}{R_\pA r - c^2 \cos\theta}  
\cH_\pA \biggr .
 \\ \biggl . 
 &\,
 +\sum_\pA \frac{(c^2 + m_- r) \scal{ m_- R_\pA ( R_\pA - r\,\cos \theta ) + c^2 ( R_\pA \cos \theta - r - m_- \sin^2 
\theta }}{m_-^{\; 2} ( R_\pA r -c^2 \cos \theta)}  \cH_\pA \biggr) d\varphi \,. \nn 
\end{align}
The latter formula is only valid in the axisymmetric case, but the generalisation to arbitrary solutions is 
straightforward, although not particularly illuminating. The Maxwell field strength is defined according to 
\eqref{eq:inst-gaug-gen}.

\subsection{Regularity of the multi-centre base}

We now consider the restrictions imposed by smoothness on the multi-centre instanton above. Note that for an actual
four-dimensional instanton, regularity would of course require the metric to be positive definite and therefore the
function $V$ in \eqref{eq:V-mult} to be strictly positive everywhere. In practical terms, this means that all the
$n_\pA<0$ in the above expressions. However, we will consider smooth five-dimensional manifolds of Minkowski signature
that admit this metric as a Euclidean four-dimensional base metric. The regularity condition on the base metric is
slightly more general, because the fibration involves an independent function that may itself also change sign.
It follows that the sign of the base metric may switch sign as long as the total five-dimensional metric remains well
defined. Physically, this change of sign corresponds to the presence of an evanescent ergo-surface, \ie an ergosurface
on which the timelike Killing vector has a double zero. The presence of such  evanescent ergo-surfaces is in fact
expected in multi-centre smooth solutions, and we will indeed find in practice that $n_\pA$ must be strictly positive
for the geometry to be smooth and globally hyperbolic in five dimensions, at least for the two-centre example we will
describe in this paper.  

In order to check regularity near the centres, we first record the poles of $V$ at the these points, as
\begin{align}
V= &\, \frac{c + m_{-}}{4 c}\, \left(c + m_{+} -  \sum_\pA\frac{2\,c\,k\,n_\pA}{ |R_\pA|+m_{-} 
}\right)\,\frac1{r-c}+\mathcal{O}((r-c)^0)\,,
\CR
V= &\, -\frac{k\,n_\pA}{r_\pA}+\mathcal{O}(r_\pA^0)\,.
\end{align}
The first of these, at the non-extremal centre, dictates the periodicity of $\psi$,
so that the geometry is free of conical singularities, if
\begin{equation}
\frac{c + m_{-}}{4 c}\, \left(c + m_{+} - 2\, \sum_\pA\frac{2\,c\,k\,n_\pA }{|R_\pA|+m_{-}}\right) = k \in 
\mathbb{R}^+\,,
\end{equation}
where $k$ defines the periodicity of the fibre coordinate as $\psi \approx \psi + 4\pi\, k$.
In general, one could set $k=1$, but we find it convenient to keep it explicitly, because it is
dimensionfull in four dimensions. One can solve this constraint by fixing 
\begin{equation}m_+ = c\, \Scal{ - 1 + \frac{4k}{c+m_-} + \sum_\pA \frac{2\, k\, n_\pA}{|R_\pA|+m_{-}}} \ .  \label{mm} \end{equation}
Similarly, the extremal centres only exhibit an
$\mathds{R}^4/\mathds{Z}_{|n_\pA|}$ orbifold singularities for $n_\pA\in \mathbb{Z}$,
which disappear for $n_\pA=\pm 1$. 

Finally, the Kaluza--Klein vector $w^0$ in \eqref{eq:w0gen} turns out to be automatically regular at the extremal 
centres up to a well defined patching condition associated to the Hopf fibration, due to the selfduality of the metric 
at these points. In contrast, the Kaluza--Klein vector generically carries an independent Dirac string singularity 
ending on the bolt, which can be resolved by setting
\begin{equation}\label{eq:Dir-bolt}
c + m_{-} - \frac{4\, c\, k}{c + m_{-}} - 2\,(c + m_{-})\,\sum_\pA\frac{k\, n_\pA}{R_\pA+m_{-}} = 2\, n\, k \,,
\end{equation}
where $n\in\mathbb{Z}$. Assuming these conditions to be satisfied, the geometry is then smooth for any integer $n$ and 
$n_\pA=\pm1$, although it is ambi-polar for positive $n_\pA$.

\subsection{Spin structure} \label{sec:spin}
A further property of the multi-centre Euclidean bases of section \ref{sec:instanton} that we will be interested in
is the existence of a spin structure on these manifolds. Recent work of \cite{Compere:2009iy,Gibbons:2013tqa} has shown that a lack of
spin structure may imply that some otherwise regular supergravity solutions violate the BPS bound, and are
therefore unphysical. These examples are in fact based on Riemannian manifold similar to the above, so we discuss
the possible obstructions to the existence of a spin structure arising from self-intersecting
homology 2-cycles. In this section we will discuss the two kinds of 2-cycles that appear in the Euclidean
Maxwell--Einstein instantons discussed in the preceding sections. Note that an obstruction to the  existence of a
spin structure is a local topological property of the associated obstruction 2-cycle supporting the 
second Stiefel--Whitney class. Therefore this discussion applies to general solutions even if we restrict our analysis 
to metrics including only the cycles of interest. 

We shall start with the $S^2$ cycle of NUT charge $n$ located at the bolt ${\rm B}$. For this purpose we
shall not consider the exact metric of the solution, but rather the simpler metric 
\begin{equation}
ds^2 = d\rho^2 + \frac{1}{4} \rho^2 \scal{ d\psi + n \cos\theta d\varphi}^2 
       +\ell^2  \scal{  d\theta^2 + \sin^2\theta d\varphi^2 }\,,
\end{equation}
that defines the same topology around the bolt, \ie a cone over $S^3/ \mathds{Z}_{|n|}$ that collapses to an
$S^2$ at the tip. Using the vielbein
\begin{equation}
e_1 = \ell d\theta \ , \quad e_2  = \ell \sin\theta d\varphi \ , 
\quad e_3 = d\rho \ , \quad e_4 = \frac{\rho}{2} \scal{ d\psi + n \cos\theta d\varphi} \ ,
\end{equation}
one computes the spin-connection $\omega$ in the neighborhood of the bolt at $\rho=0$
\begin{equation}
\omega = 
\left(\begin{array}{cccc} \ 0 \ & - \cos\theta d\varphi \ & \ 0 \ & \ 0 \\
\ \cos\theta d\varphi \ & \ 0 \ &\ 0 \ & \ 0 \ \\ 
0 \ &\ 0 \ &\  0 \ &-\tfrac{1}{2} ( d\psi + n \cos\theta d\varphi  ) \\
0&0&\ \tfrac{1}{2} ( d\psi + n \cos\theta d\varphi  ) \ & 0 \end{array}\right) + \mathcal{O}(\rho) \ .
\end{equation}
For a family of loops $\gamma_t\subset {\rm B}$, with $t\in[0,1]$, starting and ending at the same point
$x_0\in {\rm B}$ such that $\gamma_0$ and $\gamma_1$ are constant, the existence of a spin structure
requires that the $S^1$ family of holonomy loops 
\begin{equation}W(t) = \exp\Scal{ \int_{\gamma_t}\omega }\,, \end{equation}
is homotopically trivial in $SO(4)$ \cite{Hawking:1977ab}. Because the spin-connection is abelian, this is equivalent to 
the
requirement that the exponential of the integral of the Riemann curvature in the spinor representation must
be the identity. One computes 
\begin{equation}
\int_{\rm B}  R 
= \frac{i}{2} \scal{ 1-\tfrac{n}{2} \sigma_3} \otimes \sigma_3\, \int_{\rm B} \sin\theta d\theta \wedge d\varphi  
= 2\pi i \scal{ \mathds{1}- \tfrac{n}{2} \sigma_3} \otimes \sigma_3 \,, \end{equation}
and therefore
\begin{equation} \exp\Scal{ \int_{\rm B}R } = ( -1)^{n} \ .  \end{equation}
It follows that the space-time cannot admit a spin-structure if the base space includes a bolt 2-cycle
carrying an odd NUT charge. 

Let us now consider the bubble cycle linking the bolt and a Gibbons--Hawking centre. For simplicity, we shall again 
consider a simpler metric giving rise to the same topology 
\begin{equation}\label{eq:bubb-metr}
ds^2 = V_0^{-1} \Scal{ d\psi +  \tfrac{( n\, r_1 + r)\, \cos \theta - R}{r_1} d\varphi}^2 
+ V_0\, dr^2 + \Scal{ \ell + \frac{r^2}{r_1}} \scal{ d\theta^2 + \sin^2\theta d\varphi^2} \,,
\end{equation} 
with 
\begin{equation}V_0= \frac{1}{r} + \frac{1}{r_1}\,, \qquad r_1 = \sqrt{ r^2 - 2 R r \cos\theta + R^2} \ . \label{V0GH} \end{equation}
One finds indeed that this metric admits a bolt of NUT charge $n$ at $r=0$ and a self-dual nut of unit
NUT charge at $r_1=0$. Using the vielbein
\begin{equation}
\begin{split}
e_1 &= \sqrt{ \ell +  \frac{r^2}{r_1}} \, d\theta \,,\\
e_3 &= V_0^\frac{1}{2}\, dr \,,
\end{split} \qquad \begin{split}
e_2 &= \sqrt{ \ell +  \frac{r^2}{r_1}}\,  \sin\theta d\varphi \,,\\
e_4 &= V_0^{-\frac{1}{2}} \Scal{ d\psi +  \tfrac{( n\, r_1 + r)\,\cos \theta - R}{r_1} d\varphi}\,,
\end{split}\end{equation}
one computes the pull back of the spin-connection $\omega$ to the bubble $\Delta$ located at ($\theta=0$, $0\le r\le R$)
\begin{equation}
\omega|_\Delta = 
d\psi  \left(\begin{array}{cccc} \ 0 \ &\ 
\frac{r}{2R} \frac{  n ( R-r)^2 - r^2 }{r^2 + \ell (R-r )} \ & \ 0 \ & \ 0 \\
-\frac{r}{2R} \frac{  n ( R-r)^2 - r^2 }{r^2 + \ell (R-r )}  \ & \ 0 \ &\ 0 \ & \ 0 \ \\ 
0 \ &\ 0 \ &\  0 \ &\ \frac{r}{R}- \frac{1}{2}  \\
0&0&\ \frac{1}{2} -  \frac{r}{R}  \ & 0 \end{array}\right) \ . 
\end{equation}
Once again, the connection is abelian, and one can simply compute the integral of the Riemann
tensor in the spinor representation to compute the potential obstruction to the existence of
a spin structure. One obtains 
\begin{equation}
\int_\Delta  R = \int_0^{4\pi} \omega|_{r=0} - \int_0^{4\pi} \omega|_{r=R}
  = \pi i \mathds{1} \otimes \sigma_3+ 2\pi i\sigma_3 \otimes \sigma_3 \, 
\end{equation}
and therefore 
\begin{equation}\exp\Scal{ \int_{\rm \Delta}R } = -1 \ . \end{equation}
Independently of the NUT charge at the bolt, a bubble linking a Gibbons--Hawking centre to a bolt
always defines an obstruction to the existence of a spin structure. Such bubble cycles appear
generically in the multi-centre instantons discussed in section \ref{sec:instanton}, and 
space-times admitting such Euclidean bases do not admit a spin structure in general, independently
of the parity of $n$.

Even though we shall not consider this kind of solutions in this paper, let us consider as a last example the case of a 
bubble between two Gibbons--Hawking like centres. For this purpose we will consider the metric 
\begin{equation}ds^2 = V_0^{-1} \Scal{ d\psi +  \tfrac{( n\, r_1 +n_1 r)\, \cos \theta -n_1 R}{r_1} d\varphi}^2 
+ V_0\,\scal{  dr^2 + r^2  \scal{ d\theta^2 + \sin^2\theta d\varphi^2}}  \,,
\end{equation} 
with $V_0$ and $r_1$ defined as in \eqref{V0GH}. This metric reduces to a Gibbons--Hawking instanton for $n=n_1=1$, but 
it also defines a smooth geometry if they have opposite sign, \eg $n=1,\, n_1=-1$, in which case the Riemann tensor is 
not selfdual (neither is it Ricci flat). One computes for this metric 
\begin{equation}\int_\Delta  R = \int_0^{4\pi} \omega|_{r=0} - \int_0^{4\pi} \omega|_{r=R}
  = \pi i (n-n_1) \mathds{1} \otimes \sigma_3 \, 
\end{equation}
and therefore 
\begin{equation}\exp\Scal{ \int_{\rm \Delta}R } = (-1)^{n-n_1} \ . \end{equation}
Of course there is no obstruction in the case of a Gibbons--Hawking instanton, but we see that there is also no 
obstruction if both centres have opposite NUT charges.

\section{The almost-BPS system with Maxwell--Einstein base}
\label{sec:NE-sys}
In this section we define a first order system of equations that solves the equations of motion of $D=4$,
$\cN=2$ supergravity coupled to a symmetric scalar manifold and allows for non-extremal solutions.
The intuition for obtaining the system comes from the floating brane system of \cite{Bena:2009fi}, that
arose as a generalisation of the almost-BPS system \cite{Goldstein:2008fq} by replacing the hyper-K\"ahler
base of supersymmetric solutions with the metric of a solution to the Euclidean Maxwell--Einstein theory.
We therefore consider an analogous procedure in the three-dimensional non-linear sigma model obtained by timelike reduction
of the four-dimensional theory.

A stationary four-dimensional metric can be written as a timelike fibration over a three-dimensional base
\begin{equation}\label{FourDmetric} 
ds^2 = - e^{2U} (dt + \omega)^2 + e^{-2U} ds_3^2 \,,
\end{equation}
such that the electromagnetic fields take the form 
\begin{equation}A^\Lambda = \zeta^\Lambda ( dt + \omega) + w^\Lambda \ . \end{equation}
Using the equation of motions of the vectors $w^\Lambda$ and $\omega$, one can dualise them to scalars, so
that the effective three-dimensional theory reduces to a non-linear sigma model coupled to Euclidean
gravity \cite{Breitenlohner:1987dg}. 

For $\cN=2$ supergravity theories with a symmetric special K\"{a}hler target space $G_4/(U(1) \times K_5)$, the resulting
three-dimensional scalar manifold is a symmetric para-quaternionic coset space $G_3/(SL(2) \times G_4)$
\cite{Ferrara:1989ik}. One can then consider a coset representative, $\cV$, and define 
the scalar momenta $P_i$ as the coset component of the Maurer--Cartan form 
\begin{equation}d\cV \, \cV^{-1} = P + B  \  \end{equation}
in $\mathfrak{g}_3 \ominus (\mathfrak{sl}_2 \oplus \mathfrak{g}_4)$. The equations of motion can then be cast in the 
conservation of the $\mathfrak{g}_3$ current for the scalars and the three-dimensional Einstein equation, as
\begin{equation}\label{eq:3Deoms}
d \star \scal{ \cV^{-1} P \cV }= 0 \ , \qquad 
R_{ij} = \mbox{tr} \  P_i P_j \  .
\end{equation}

A powerful method for obtaining solutions arises by assuming that $\cV$ lies in a nilpotent subgroup of $G_3$.
Then $P_i$ is automatically nilpotent such that the three-dimensional base metric is necessarily flat, because
then $R_{ij}=0$, and the system of equations for the scalars is solvable. This guarantees the existence of multi-centre
solutions, which can be used for example to obtain either supersymmetric or non-supersymmetric black holes or smooth
five-dimensional geometries. Such solvable systems are classified by the complex $SL(2)\times G_4$ orbits of the generic
asymptotic value of $P_i$ in the system, and one can therefore refer to the classification of nilpotent orbits of $G_3$
to classify all the independent such solvable systems \cite{Bossard:2011kz,Bossard:2012ge}. However, the corresponding
solutions are by construction extremal, because the three-dimensional base is necessarily flat. 

This eliminates for instance non-extremal solutions as well as extremal solutions of Kerr type, for which the BPS
bound is not saturated, and the asymptotic Noether charge 
\begin{equation}Q = \frac{1}{4\pi} \int_{S^2_{\infty}} \cV^{-1} P \cV \label{G3Charge} \end{equation}
is not nilpotent. For a black hole solution, non-extremality is defined as having a non-zero Hawking temperature, 
or equivalently a non-zero surface gravity on the horizon. For a smooth solution there is no horizon, and there is no
such geometric criterion to define extremality. Algebraically, one can define the extremality parameter $c$ such that
the asymptotic charge $Q$ in the fundamental representation satisfies to \cite{Bossard:2009at} \footnote{For $E_8$ type
groups, there is an equivalent quintic equation in the {\bf 3875}.} 
\begin{equation}Q^3 = c^2 \, Q \ \label{Kerr-Orbit} \end{equation}
for a regular single-centre four-dimensional black hole, where $c$ is the parameter appearing in the three-dimensional 
metric
\begin{equation}\label{eq:Kerr-base}
ds_3^{\; 2} =   \Scal{ 1  - \frac{ a^2 \sin^2 \theta}{r^2 - c^2 + a^2 } } dr^2 + \scal{ r^2 - c^2 + a^2 
\cos^2\theta} d\theta^2 +  \scal{ r^2 - c^2 + a^2 } \sin^2\theta d\varphi^2 \,,
\end{equation}
in spherical coordinates. Note that for the extremal Kerr solution $c=a\ne0$, but this will not be important in this 
paper because we will only consider solutions with $a=0$. The Euclidean Maxwell--Einstein instantons defined in the 
preceding section have indeed this three-dimensional base metric for  $a=0$, as do static single-centre black holes. 
What we will mean by a smooth non-extremal solution in this paper is a smooth solution for which the asymptotic charge 
$Q$ satisfies to equation \eqref{Kerr-Orbit} for a strictly non-vanishing $c$. Note that
for a four-dimensional stationary space-time, the three-dimensional base metric is uniquely defined, but it is
ambiguous for five-dimensional doubly axisymmetric stationary space-times, depending of the choice of
isometry. Nonetheless, the five-dimensional solutions we will consider in this paper all asymptote locally to 
$S^1 \times \mathds{R}^{1,3}$, so that the isometry is naturally defined to be the one acting on the finite size
$S^1$ factor. 

In section \ref{sec:sys-def} we consider a deformation of the almost-BPS system in the three-dimensional non-linear
sigma model, which includes a subsystem solving the Euclidean Maxwell--Einstein equations as in \cite{Bena:2009fi},
and allowing for a non-flat three-dimensional base metric as \eqref{eq:Kerr-base}. We then analyse the asymptotic
properties of the solutions of the particular system we
consider in section \ref{sec:BPSbound}, where we show that they admit necessarily asymptotic scalar momenta inconsistent with the regular black hole attractor flow. For completeness, we discuss the possible extremal
limits of the system in section \ref{sec:extremal}, where we show that all known extremal systems, namely the BPS
\cite{Behrndt:1997ny, Denef:2000nb}, almost-BPS \cite{Goldstein:2008fq} and composite non-BPS \cite{Bossard:2011kz}
systems can be obtained in this way. Finally, in section \ref{sec:5D-lift} we connect to the floating brane system
\cite{Bena:2009fi}  by lifting to five-dimensional supergravity, where it was originally obtained.

\subsection{Definition of the system}\label{sec:sys-def}

For the asymptotic charge to satisfy to a characteristic equation like \eqref{Kerr-Orbit} while keeping the system 
partially solvable, we shall consider $\cV$ in a relevant parabolic subgroup of $G_3$. For example, the almost BPS 
system in the exceptional $\cN=2$ supergravity theory with moduli parametrising the special K\"{a}hler symmetric space 
$E_{7(-25)}/ ( U(1) \times E_{6(-78)})$ \cite{Gunaydin:1983bi}, is associated to the graded decomposition of $\mathfrak{g}_3 \cong 
\mathfrak{e}_{8(-24)}$, as
\begin{multline}\mathfrak{e}_{8(-24)} \cong {{\bf 1}^\ord{-5}} \oplus  {\bf 1}^\ord{-4}\oplus {{\bf 27}^\ord{-3}} \oplus 
\overline{\bf 27}^\ord{-2}  \oplus {( {\bf 1}\oplus \overline{\bf 27})^\ord{-1} } \\  \oplus \scal{ \gl_1  \oplus 
\gl_1\oplus  \mathfrak{e}_{6(-26)} }^\ord{0}  \oplus {( {\bf 1}\oplus {\bf 27})^\ord{1} }\oplus {\bf 27}^\ord{2} \oplus 
{\overline{\bf 27}^\ord{3}} \oplus {\bf 1}^\ord{4}\oplus {{\bf 1}^\ord{5}} \,,
\end{multline} 
where the generators of $\cV$ are restricted to the odd positive grade elements, \ie $V , {\bf K}$ at grade 1, ${\bf L}$
at grade 3 and a function $M$ at grade 5. To get a non-trivial three-dimensional metric, one must necessarily turn on
some of the negative grade generators. The simplest solution is to turn on the grade $-1$ elements
$\bar V, \bar {\bf K} \in {\bf 1}\oplus \overline{\bf 27}$, but one must then constrain $\bar {\bf K}$ in order to
avoid mixing with all the fields of the theory and keep a solvable structure.

The relevant condition turns out to be $\bar {\bf K} \times \bar {\bf K} = 0 $ in the ${\bf 27}$, where the cross
product is defined from the cubic $E_{6(-26)}$ invariant, \ie the  $c_{ijk}$ symmetric tensor defining the
prepotential for cubic models in general. This suggests the further decomposition of $ \mathfrak{e}_{6(-26)}$ into 
\begin{equation} 
\mathfrak{e}_{6(-26)} \cong 
\overline{\bf 16}^\ord{-3} \oplus \scal{ \gl_1 \oplus \so(1,9)}^\ord{0} \oplus {\bf 16}^\ord{3}
\end{equation}
with respect to which 
\begin{equation}
{\bf 27} \cong {\bf 1}^\ord{-4} \oplus {\bf 16}^\ord{-1} \oplus {\bf 10}^\ord{2} \ . 
\end{equation}
Note that this decomposition corresponds to the duality group of the theory in six dimensions. This decomposition holds 
in general for theories lifting to six dimensions, so that ${\bf 16}$ corresponds in general to the representation of 
the vector multiplets, and ${\bf 10}$ to the representation of the tensor multiplets. Accordingly, we consider a generalised Ansatz for $\cV$, that includes in addition to the fields of the standard almost BPS system, both $\bar V$ and the function $\bar K$ in the ${\bf 1}^\ord{4}$ component of $\overline{\bf 27}$. The system will exhibit a manifest $Spin(1,9)$ symmetry, which in other theories would similarly be the duality symmetry of the six-dimensional uplift. The four
distinguished functions, $V,\bar V, \bar K$ and $K$ in the ${\bf 1}^\ord{-4}$ component of ${\bf 27}$ parametrise
altogether a $SL(3)/GL(2)$ coset space that corresponds to a subsystem solving axisymmetric Euclidean Maxwell--Einstein equations. 
It is therefore convenient to reorganise the decomposition of $\mathfrak{e}_{8(-24)}$ so that the $\sl_3$ is in the
grade zero component, as
\begin{multline} 
\mathfrak{e}_{8(-24)}  \cong   \overline{\bf 3}^\ord{-4}\oplus {\bf 16}^\ord{-3}\oplus({\bf 3}\otimes 
{\bf 10})^\ord{-2}  \oplus (\overline{\bf 3}\otimes \overline{\bf 16})^\ord{-1}\\ \oplus  \scal{ \gl_1  \oplus 
\mathfrak{sl}_3\oplus  \mathfrak{so}(1,9) }^\ord{0}  \oplus ({\bf 3}\otimes {\bf 16})^\ord{1} \oplus(\overline{\bf 
3}\otimes {\bf 10})^\ord{2}\oplus \overline{\bf 16}^\ord{3} \oplus  {\bf 3}^\ord{4}\ .\label{ParaGrad}  
\end{multline}

The system we shall consider in this paper consists in restricting $\cV$ to the parabolic subgroup defined as the 
semi-product of the $SL(3)$ grade zero component with the positive grade nilpotent subgroup, as
\begin{equation}\cV \in SL(3) \ltimes \mathds{R}^{3\times 16+3\times 10+16+3} \ . \end{equation}
To exhibit the independent functions defining the system, we must also consider the corresponding decomposition of the 
divisor subgroup $SL(2) \times E_{7(-25)}$, \ie
\begin{eqnarray}
 \sl_2   &\cong&  {\bf 1}^\ord{-4} \oplus   \gl_1{}^\ord{0} \oplus  {\bf 1}^\ord{4} \CR
 \mathfrak{e}_{7(-25)}  &\cong&     {\bf 10}^\ord{-2} \oplus({\bf 2}\otimes \overline{\bf 16})^\ord{-1}\oplus   \scal{  
\gl_1  \oplus \mathfrak{sl}_2\oplus  \mathfrak{so}(1,9) }^\ord{0}  \oplus ({\bf 2}\otimes {\bf 16})^\ord{1}\oplus   {\bf 
10}^\ord{2}\,. \label{ParaGradK}
\end{eqnarray}
According to this decomposition, the nilpotent subgroup is parametrised by the $54$ real functions 
\begin{equation}
K^\alpha \in {\bf 16}^\ord{1} \ , \quad ( K^a,\, L^a) \in ({\bf 2}\otimes {\bf 10})^\ord{2} \ , \quad L_\alpha \in 
\overline{\bf 16}^\ord{3} \ , \quad (L,M) \in {\bf 2}^\ord{4}\ .
\end{equation}
The system can then admit a non-trivial three-dimensional metric through its coupling to the $SL(3)/GL(2)$ non-linear 
sigma model. Therefore, starting with a given axisymmetric solution of
Euclidean Maxwell--Einstein equations, we get a solvable system of differential equations
for $ K^\alpha,\, K^a,\, L_a,\, L_\alpha,\, L,\, M$. 

By solvable we mean that the grade 1 functions $K^\alpha$ satisfy to linear differential equations 
that depend on the specific Euclidean electro-vacuum we start with, the grade 2 functions $K^a,\, L_a$  satisfy to 
linear equations with source terms depending quadratically on $K^\alpha$ and its derivative, and similarly the higher 
grade functions $L_\alpha,\, L,\, M$ satisfy to linear equations with sources that depend polynomially on the lower
grade functions $K^\alpha,\, K^a,\, L^a$ and their derivatives. The structure of the system is therefore similar to
the one associated to extremal solutions, the difference being that instead of having simply Poisson equations on
$\mathds{R}^3$, one gets more complicated linear equations that depend non-trivially on the chosen Euclidean
electro-vacuum solution. Within the STU truncation, the only non-trivial functions are then
$K^a = ( K^2,K^3),\, L=L_1,\, L_a = ( L_2,L_3),\, M$ and this system of equations is equivalent to the one
derived from the floating brane Ansatz in \cite{Bena:2009fi}. 

There is a different system that can be obtained by decomposing $\mathfrak{e}_{8(-24)}$ in the same way as in
\eqref{ParaGrad}, but with a different graded decomposition of the divisor subgroup such that the $SL(2)$ component does 
not decompose and 
\begin{multline} 
\mathfrak{e}_{7(-25)}  \cong  {\bf1}^\ord{-4} \oplus  {\bf 16}^\ord{-3}\oplus   {\bf 10}^\ord{-2} \oplus 
\overline{\bf 16}^\ord{-1} \\
\oplus   \scal{  \gl_1  \oplus \mathfrak{gl}_1\oplus  \mathfrak{so}(1,9) }^\ord{0}  \oplus 
{\bf 16}^\ord{1}\oplus   {\bf 10}^\ord{2} \oplus  \overline{\bf 16}^\ord{3} \oplus {\bf 1}^\ord{4} \,,
\end{multline} 
In  this case the nilpotent subgroup is parametrised by the $54$ real functions 
\begin{equation}( K^\alpha ,\, L^\alpha) \in ({\bf 2}\otimes {\bf 16})^\ord{1} \ , \quad ( K^a,\, L^a) \in ({\bf 2}\otimes {\bf 
10})^\ord{2} \ , \quad (V,M) \in {\bf 2}^\ord{4}\ . \end{equation}
Note however that both systems would be identical within $\cN=8$ supergravity. In the same way, the relevant graded 
decomposition of $\e_{8(8)}$ is associated to the six-dimensional duality group 
\begin{multline}\label{E8grad}  
\mathfrak{e}_{8(8)}  \cong   \overline{\bf 3}^\ord{-4}\oplus {\bf 
16}^\ord{-3}\oplus({\bf 3}\otimes {\bf 10})^\ord{-2}  \oplus (\overline{\bf 3}\otimes \overline{\bf 16})^\ord{-1}\\ 
\oplus  \scal{ \gl_1  \oplus \mathfrak{sl}_3\oplus  \mathfrak{so}(5,5) }^\ord{0}  \oplus ({\bf 3}\otimes {\bf 
16})^\ord{1} \oplus(\overline{\bf 3}\otimes {\bf 10})^\ord{2}\oplus \overline{\bf 16}^\ord{3} \oplus  {\bf 3}^\ord{4}\,, 
\end{multline} 
with the graded decomposition of the $Spin^*(16)$ subgroup 
\begin{multline}\label{so16grad} 
\mathfrak{so}^*(16) \cong  {\bf 1}^\ord{-4}\oplus ({\bf 2}_2\otimes \overline{\bf 
4})^\ord{-3}\oplus({\bf 6} \oplus {\bf 2}\otimes {\bf 2}_1\otimes {\bf 2}_2)^\ord{-2}  \oplus ({\bf 2}\otimes{\bf 2}_1 
\otimes \overline{\bf 4}\oplus {\bf 2}_2 \otimes {\bf 4})^\ord{-1}\\ \oplus  \scal{ \gl_1  \oplus \gl_1 \oplus 
\mathfrak{sl}_2\oplus \su(2)_1 \oplus \su(2)_2 \oplus  \mathfrak{su}^*(4) }^\ord{0}  \\
\oplus ({\bf 2}\otimes{\bf 2}_1 \otimes {\bf 4}\oplus {\bf 2}_2 \otimes \overline{\bf 4})^\ord{1}\oplus({\bf 6} \oplus 
{\bf 2}\otimes {\bf 2}_1\otimes {\bf 2}_2)^\ord{2}\oplus ({\bf 2}_2\otimes{\bf 4})^\ord{3}\oplus  {\bf 1}^\ord{4} \,,
\end{multline} 
In this case we get $50$ real functions 
\begin{eqnarray} 
( K_{\alpha_1}^a ,\, K^{\alpha_2}_a,\, L^{\alpha_2}_a) &\in& 
( {\bf 2}_1 \otimes {\bf 4} \oplus {\bf 2} \otimes {\bf 2}_2\otimes \overline{\bf 4})^\ord{1} \,,
\CR
(K_{ab} ,\, L_{ab},\, Y^{\alpha_2}_{\alpha_1}) &\in& 
( {\bf 2}\otimes {\bf 6} \oplus {\bf 2}_1\otimes {\bf 2}_2)^\ord{2} \,,
\CR
L_{\alpha_1 a} &\in& ( {\bf 2}_1 \otimes \overline{\bf 4})^\ord{3} \,,
\CR
(L,\, M) &\in& {\bf 2}^\ord{4}\,,
\label{N8Functions} 
\end{eqnarray}
on top of the four functions parametrising the Euclidean electro-vacuum. This system admits then two inequivalent
truncations to $\cN=2$ supergravity defined by eliminating the fields transforming non-trivially with respect to one of the two $SU(2)$ automorphisms of the system, giving rise to the two different systems we have discussed in this section. 

In this paper we will restrict our analysis to the system described by the decomposition 
\eqref{ParaGrad}-\eqref{ParaGradK} when the spinor fields $K^\alpha, \, L_\alpha$ are set to zero. This corresponds 
effectively to restricting ourselves to the axion dilaton models with special K\"{a}hler space 
$SL(2)/SO(2) \times SO(2,n)/(SO(2)\times SO(n))$. The system of differential equations then reduces essentially to
the floating brane system of \cite{Bena:2009fi}. 
Using an appropriate explicit representation of this parabolic subgroup in $SO(4,4)$, one computes that the associated 
Ansatz is indeed of the standard form for the metric, with
\begin{equation}\label{eq:eU}
e^{-4U} = \tfrac{1}{2}\,V \, L \, L_a L^a  - M^2 \ , \qquad
     V^{-1} = \cE_+  + \cE_- + K_+  K_- \ ,
\end{equation}
 and
\begin{equation}\label{eq:do}
\star d\omega =  d M - V\, L d K_+  - V  \, L_a dK^a 
 + 2 \,V\, M \,(d \cE_+ + K_- d K_+) + V\,   L_a L^a\, d K_- \ ,
\end{equation}  
where vector indices $a$ are raised with the $SO(1,n-1)$ metric $\eta^{ab}$. For the STU model, which corresponds
to $n=2$, this matrix is 
\begin{equation}\label{eq:STU-eta}
\eta^{ab}  = \left(\begin{array}{cc} 0\ &\ 1\\1\ &\ 0\end{array}\right) \,,
\end{equation}
and we have then the identification $L=L_1,\, L_a = (L_2,L_3)$, such that $\tfrac{1}{2} L L_a L^a = L_1 L_2 L_3$, and 
$K_+ = K^1 ,\, K^a = (K^2 ,K^3)$.

The scalar fields are defined as
\begin{equation}t^1= K_+ + \frac{-M + i e^{-2U} }{V L} \ , \qquad t^a = K^a +2 L^a \frac{-M + i e^{-2U} }{V L_b L^b }  \ .  \end{equation}
The electric vector field components are
\begin{eqnarray} \label{eq:alm-NE-mag}
\star d w^0 &=&  V^2\, (d \cE_+ - d \cE_- + K_- d K_+ - K_+ d K_-)\,, 
\CR
\star d w^a &=& V\, d K^a - K^a dw^0 - 2\,V\, L^a \, d K_- \,, 
\CR
\star d w^1 &=& V\, d K_+ - K_+ dw^0\ ,
\end{eqnarray}
while the magnetic field components are
\begin{eqnarray}  \label{eq:alm-NE-el}
\star d v_a &=& d L_a +  K_a K_+ dw^0 -  V\, d (K_a K_+) 
+ 2\,V\, L_a K_+ \, d K_- \,, 
\CR
\star d v_1 &=& d L +  \tfrac{1}{2} K_a K^a  dw^0 - \tfrac{1}{2}  V\,d (K^a K_a) 
- 2 d(M K_-) -  4\,V\, M K_- \,(d \cE_+ + K_- d K_+) \,,
\CR
&& + 2\,V\, L  K_- dK_+  + 2\,V\, L_a d(K_- K^a ) - 2\,V\,  L_a L^a K_- \, d K_- \,,
\CR
\star d v_0 &=& \tfrac{1}{2}  K^a K_a K_+ dw^0 - \tfrac{1}{2} V\, d( K^a K_a K_+)
+ K_+\,d L + K^a dL_a   + 2 \,V\, K^a L_a K_+\, d K_- \,,
\CR
&& -
 2\, K_+ d(M K_-)  - 2 \cE_+ \,d M -
 2\,V\, (  \cE_+ - \cE_- + K_- K_+)\, M \,(d \cE_+ + K_- d K_+)   \CR
&& +V\, (   \cE_+ - \cE_- +K_- K_+) (L\,d K_+ + L_a dK^a  - L_a L^a \, d K_-)   \,.
 \end{eqnarray}
The equations of motion then follow from the Bianchi identity of these vector fields. The functions $\cE_\pm$ and $K_\pm$ are then identified with the corresponding solution of  Euclidean
Maxwell--Einstein equations of the previous section, so they are solutions to \eqref{eq:EK-eoms}, while
the three-dimensional metric is given by \eqref{eq:R-base}.

\subsection{Asymptotic structure and the BPS bound}
\label{sec:BPSbound}
Using the general system based on \eqref{E8grad}-\eqref{N8Functions} in $\cN=8$ supergravity
for defining non-extremal solutions, one can already make a general comment
on the property of the total charge \eqref{G3Charge}. Let us define the asymptotic momentum
$P_\infty = \cV_\infty Q \cV^{-1}_\infty$, which lies by definition in the coset component of 
the parabolic sub-algebra, \ie 
\begin{equation}
P_\infty \in  ( {\bf 2} \oplus {\bf 2})^\ord{0} \oplus ( {\bf 2}_1 \otimes {\bf 4} \oplus {\bf 2} \otimes 
{\bf 2}_2\otimes \overline{\bf 4})^\ord{1}\oplus  ( {\bf 2}\otimes {\bf 6} \oplus {\bf 2}_1\otimes {\bf 2}_2)^\ord{2} 
\oplus  ( {\bf 2}_1 \otimes \overline{\bf 4})^\ord{3} \oplus  {\bf 2}^\ord{4}  \ . 
\label{Pin} 
\end{equation}
 If the solution admits the total charge $Q$ of a non-extremal black hole,
it must by construction satisfies \eqref{Kerr-Orbit}, and so does $P_\infty$. This implies that $P_\infty$
is diagonalisable, and therefore one can rotate it to the grade zero component in \eqref{E8grad} and \eqref{Pin}.
It follows that $P_\infty$ lies in the $Spin^*(16)$ orbit of a generic element of the coset component of $\sl_3$ in
\eqref{E8grad}. One straightforwardly computes its stabilizer in $\so^*(16)$ within the decomposition
\eqref{so16grad} as
\begin{multline}
\mathfrak{su}^*(8) \cong ({\bf 2}_2\otimes \overline{\bf 4})^\ord{-3}\oplus({\bf 2}_1\otimes {\bf 
2}_2)^\ord{-2}  \oplus ({\bf 2}_1 \otimes \overline{\bf 4})^\ord{-1}\\ \oplus  \scal{ \gl_1  \oplus \gl_1 \oplus 
\su(2)_1 \oplus \su(2)_2 \oplus  \mathfrak{su}^*(4) }^\ord{0}  
\oplus ({\bf 2}_1 \otimes {\bf 4})^\ord{1}\oplus( {\bf 2}_1\otimes {\bf 2}_2)^\ord{2}\oplus ({\bf 2}_2\otimes{\bf 
4})^\ord{3} \,,
\end{multline} 
It is clear that the stabilizer cannot be $SU(8)$, and one finds indeed that $SU^*(8)$ is the only real
form admitting this graded decomposition.

In contrast, regular black holes are necessarily in the $Spin^*(16)$ orbit of the Kerr solution \cite{Breitenlohner:1987dg}, which is
isomorphic to $Spin^*(16)/ SU(8)$ and not $Spin^*(16)/ SU^*(8)$. It follows that the above system does not
contain standard non-extremal black holes, and in fact the generic solution to this system will not have
the asymptotic structure of a regular black hole solution. However, that does not exclude the possibility
of obtaining physically relevant solutions, especially if one is interested in constructing solutions that
lift to smooth higher dimensional geometries.

In order to appreciate the implications of the appearance of the $SU^*(8)$ stabilizer, it is instructive to
consider the various extremal limits of regular black holes. The general non-extremal black hole
solution \cite{Chow:2014cca} admits several extremal limits, depending on the sign of the moduli independent
expression 
\begin{equation}\lozenge \equiv  I_4(q,p) + J^2 \ , \end{equation}
which is proportional to the product of the outer and inner horizon areas
\begin{equation}A_+ A_- = (4\pi)^2 \bigl|  \lozenge \bigr| \ . \end{equation}
Depending on the sign of $\lozenge$, 
there are two branches of extremal solutions one can obtain by taking appropriate limits of the non-extremal
solution.

In the standard branch with $\lozenge>0$,  the general solution \cite{Chong:2004na} admits two extremal limits,
the over-rotating extremal limit for which $P_\infty$ is in the $Spin^*(16)$ orbit (of stabilizer $SU(8)$) of
the extremal Kerr solution, and the BPS limit for which $J=0$ and the ADM mass saturates the BPS bound 
\begin{equation}M_{\scriptscriptstyle \rm ADM} = |Z(q,p)| \ , \end{equation}
where $Z(q,p)$ stands for the eigenvalue of the asymptotic $\cN=8$ central charge matrix $Z(q,p)_{ij}$ with the
largest modulus. Within the STU model, it would be the greatest of the four $|Z(q,p)|,\,|D_i Z(q,p)|$. In this
limit, $P_\infty$ is nilpotent and lies in the generic BPS orbit \cite{Bossard:2009at}
\begin{equation}
P_\infty \in Spin^*(16) \big/ \scal{ SU(2)\times SU(6) \ltimes ( \mathds{C}^{2\times6}\oplus \mathds{R}) }\ . 
\end{equation}
Beyond the BPS bound $M_{\scriptscriptstyle \rm ADM} < |Z(q,p)|$, the formal solution admits a diagonalisable
asymptotic momentum of stabilizer $SU(2,6) \in Spin^*(16)$.

The second branch, with $\lozenge<0$, admits a single extremal limit \cite{Rasheed:1995zv,Larsen:1999pp} in
which the asymptotic $P_\infty$ is nilpotent and lies in the non-BPS extremal orbit \cite{Bossard:2009at}
\begin{equation}P_\infty \in Spin^*(16) \big/ \scal{ Sp(4) \ltimes \mathds{R}^{27} }\ . \end{equation}
In this case the mass is determined by the fake superpotential
\cite{Ceresole:2007wx,Andrianopoli:2007gt,LopesCardoso:2007ky,Perz:2008kh,Bossard:2009we,Ceresole:2009vp}  as
\begin{equation}M_{\scriptscriptstyle \rm ADM} =  W(q,p) \ , \end{equation}
which is ensured to respect the BPS bound for a strictly negative $I_4(q,p)$. Beyond this wall in the space of
formal black hole solutions, the asymptotic momentum is diagonalisable with stabilizer $SU^*(8)$. This orbit
includes for example under-rotating over-extremal Kaluza--Klein black holes in five dimensions. These carry
electromagnetic charges with a strictly negative quartic invariant and violate the bound that their
mass should be greater than the fake superpotential, \ie
\begin{equation}M_{\scriptscriptstyle \rm ADM} < W(q,p) \ . \end{equation}
Note that for a charge of negative quartic invariant, the condition $W(q,p)>|Z(q,p)|$ still leaves room for
the BPS bound to be satisfied, but even when it is, such solutions do not correspond to physically acceptable
macroscopic black holes for which there exists a regular four-dimensional solution.

However they are not the only solutions with an asymptotic momentum $P_\infty$ of stabilizer $SU^*(8) \subset Spin^*(16)$. As we shall discuss now, there are two other branches of such solutions, which can satisfy to the regularity bound $M_{\scriptscriptstyle \rm ADM}\ge W(q,p)$, but for which the scalar flow is off the black hole attractor trajectory. To illustrate this, let us restrict ourselves to a
solution of five-dimensional pure gravity with dilaton
$e^{-2\phi} =  e^{2U} V \scal{\tfrac{1}{2} L L_a L^a}^{\frac{1}{3}}$ in our conventions. The condition
\eqref{Kerr-Orbit} then implies the cubic equation \footnote{Where $Z= \frac{1}{2} (e^{-3\phi}Q_0-i e^{3\phi}P^0)$.}\begin{equation}\dot{\phi}^3 =\scal{   M_{\scriptscriptstyle \rm ADM}^{\; 2} + 2 |Z|^2 } \dot{\phi} - M_{\scriptscriptstyle \rm ADM} \scal{ Z^2 + \bar Z^2} \ , \label{CubicLarsen} \end{equation}
where $\dot{\phi}$ is the asymptotic momentum of the dilaton, with
\begin{equation}c^2 =  M_{\scriptscriptstyle \rm ADM}^{\; 2} + 3 \dot{\phi}^2 - 4 |Z|^2 \ . \label{C2Larsen} \end{equation}
This equation is understood to determine the asymptotic momentum of the dilaton in function of the ADM mass
and the asymptotic charges, in agreement with the no-hair theorem. However, because this equation is
cubic, it clearly admits three separated branches in general, and all turn out to be real. For simplicity we will consider the solutions in the two extreme cases, when the charges vanish and when the extremality bound is saturated. 

For the case of
vanishing charge, $Z=0$, one can clearly see from \eqref{CubicLarsen} that the non-extremal black hole is obtained for
$ \dot{\phi}_{\scriptscriptstyle \rm BH}=0$, whereas the two other branches, 
$ \dot{\phi}=\pm M_{\scriptscriptstyle \rm ADM}$,  correspond to naked singularities in four-dimensions,
but lift to a smooth Euclidean Schwarzschild solution in five dimensions \cite{Bena:2009qv}. One computes that
the stabilizer of the asymptotic momentum is $SU(8)$ for the black hole solution, whereas it is $SU^*(8)$ for
the two others. These two other solutions are indeed part of the solutions that exist within the partially
solvable system we discuss in this paper.

When the ADM mass is equal to the fake superpotential $W$, such that the associated black
hole solution is extremal, the cubic equation \eqref{CubicLarsen} simplifies to 
\begin{equation}\scal{  \dot{\phi} -  \dot{\phi}_{\rm \scriptscriptstyle BH}}\Scal{  \dot{\phi} +\tfrac{1}{2}  \dot{\phi}_{\rm \scriptscriptstyle BH} +\tfrac{1}{2} \sqrt{ 6W^2 + 3  \dot{\phi}_{\rm \scriptscriptstyle BH}^{\; 2}  }}\Scal{  \dot{\phi} +\tfrac{1}{2}  \dot{\phi}_{\rm \scriptscriptstyle BH} -\tfrac{1}{2} \sqrt{ 6W^2 + 3  \dot{\phi}_{\rm \scriptscriptstyle BH}^{\; 2}  }} = 0 \end{equation}
where $ \dot{\phi}_{\rm \scriptscriptstyle BH}$ is the asymptotic momentum of the regular black hole solution
obtained as the derivative of the fake superpotential. One then finds that if the black hole solution
$ \dot{\phi}= \dot{\phi}_{\rm \scriptscriptstyle BH}$ gives by construction $c=0$, the extremality parameter
$c$ remains strictly positive for the two other solutions. In fact, using the extremality condition $c=0$ to
solve for $ \dot{\phi} $ in \eqref{CubicLarsen}, one gets by consistency a cubic polynomial for the expression
of the mass $M_{\scriptscriptstyle \rm ADM}=W$
\begin{equation}W^6  -3 |Z|^2  W^4  + \frac{3}{16} \scal{ 9 (Z^2 + \bar Z^2)^2 - 20 |Z|^4} W^2 - 
 | Z|^6 = 0 \ .\label{FakePot} \end{equation}
It turns out that the only real root is the fake superpotential $W$ if we assume $W> |Z|$, and we conclude that the two
other branches do not include extremal solutions and generally give rise to an asymptotic momentum of
stabilizer $SU^*(8)$. 

%
\begin{table}[t]
\centering
{\renewcommand{\arraystretch}{1.7}
\begin{tabular}{|c|c|c|}
\hline
\multicolumn{3}{|c|}{non-extremal}
\\ \hline
 \multicolumn{2}{|c|}{BH asympt.} & non-BH asympt. \\ \hline
 $M>W$ & $M<W$ &  $M \gtrless W$  \\ \hline
 $SU(8)$ & $SU^*\!(8)$ & $SU^*\!(8)$ 
\\ \hline
\end{tabular}
}
\caption{A summary of the stabilizers for various non-extremal solutions, where we distinguish
between solutions with the same asymptotics as regular black holes and solutions with scalar
flows that cannot describe regular black holes. The system we present in the text includes all
three branches with $SU^*\!(8)$ stabilizer.}
\end{table}

The different branches can only connect when $Z$ is real, \ie  $I_4 =   ( Z^2-\bar Z^2)^2=0$, in which case the extremality bound coincides with the BPS bound $W =|Z|$, and \eqref{FakePot} reduces to $(W^2-Z^2)^3 =0$. More generally if $Z$ is real, the roots of \eqref{CubicLarsen} reduce to 
\begin{equation}\dot{\phi} = \frac{Z M_{\scriptscriptstyle \rm ADM}}{|Z|} \ , \quad  \dot{\phi} = \frac{Z}{2|Z|} \Scal{ -M_{\scriptscriptstyle \rm ADM}  \pm\sqrt{ M_{\scriptscriptstyle \rm ADM}^{\; 2} + 8 Z^2 } } \ , \label{SmallAtrra}\end{equation}
and the first solution also tends to the extremal limit $c=0$ as $M_{\scriptscriptstyle \rm ADM}$ reaches the BPS bound.
The solution corresponding to a regular black hole is the one with the plus sign in the second of \eqref{SmallAtrra}. Both solutions coincide in the extremal limit $M_{\scriptscriptstyle \rm ADM} =|Z|$, and then  belong to an orbit of extremal black holes with vanishing horizon area. 

We know the system we consider describes solutions with $SU^*(8)$ stabilizer, which includes all three branches,
namely the over-extremal continuation of regular black hole solutions and two branches that may respect the BPS bound
but always lead to a scalar flow that is not the one of a regular black hole solution. Therefore, the
possibility of obtaining solutions that lift to smooth geometries can be realised, provided that one does not
insist on the scalar flow being the same as that of a regular black hole solution. This would signal
a deviation from the standard lore based on the construction of extremal microstate geometries, which only deviate from
the corresponding black hole solution in the region close to the location of the horizon. If solutions with
different scalar flow were to be considered admissible, the quantum state describing the black hole background
would be a superposition of pure states approximated by geometries that deviate from the original black hole
solution throughout the flow to infinity, as
\begin{equation}
|{\rm BH}\rangle = \int  d\mu \Psi_1 |{\rm Micro}\rangle_1 + \int  d\mu \Psi_2 |{\rm Micro}\rangle_2 \ , 
\end{equation}
where $ |{\rm Micro}\rangle_1$ and $|{\rm Micro}\rangle_2$ would respectively be well approximated by globally hyperbolic smooth solutions with asymptotic scalar momenta in the two other branches of solutions to \eqref{C2Larsen}. In this framework, the expectation value of the scalar asymptotic momentum better be the one of the regular black hole solution
\begin{equation}\langle {\rm BH}| \hat{\dot{\Phi} }| {\rm BH} \rangle  = \int  d\mu \Psi_1 \Psi_1^*  \, \dot{\phi}_1 + \int  d\mu \Psi_2 \Psi_2^* \,  \dot{\phi}_2 = \dot{\phi}_{\rm \scriptscriptstyle BH}\ . \end{equation}
Let us test if this is at least possible, using the simple example of a black hole of vanishing quartic invariant, \ie $Z=\bar Z$. Assuming that the black hole quantum state has the probability $ \int  d\mu \Psi_1 \Psi_1^*=x$ to be a microstate in one branch and $ \int  d\mu \Psi_2 \Psi_2^*=1-x$ to be a microstate in the other, one requires from \eqref{SmallAtrra} that 
\begin{equation}x M_{\scriptscriptstyle \rm ADM} + (1-x) \frac{1}{2} \scal{ -M_{\scriptscriptstyle \rm ADM}-\sqrt{M_{\scriptscriptstyle \rm ADM}^{\; 2} + 8 Z^2}} =  \frac{1}{2} \scal{ -M_{\scriptscriptstyle \rm ADM}+\sqrt{M_{\scriptscriptstyle \rm ADM}^{\; 2} + 8 Z^2}} \ , \end{equation}
with $0\le x\le 1$, \ie
\begin{equation}0\le \frac{2 \sqrt{M_{\scriptscriptstyle \rm ADM}^{\; 2} + 8 Z^2}}{3M_{\scriptscriptstyle \rm ADM}+\sqrt{M_{\scriptscriptstyle \rm ADM}^{\; 2} + 8 Z^2}} \le 1\ , \end{equation}
which is true if and only if $ M_{\scriptscriptstyle \rm ADM} \ge |Z|$. It is rather suggestive that the proposal is consistent if and only if the black hole mass satisfies to the BPS bound. More generally, using the parametrisation of a regular solution of \cite{Larsen:1999pp}, one can check that the same is always possible for a regular black hole solution satisfying the extremality bound $M_{\scriptscriptstyle \rm ADM} \ge W$. Namely, the solution, $x$, to
\begin{equation}x \frac{ p - q - \sqrt{-32 m^2 + 9 p^2 + 6 p q + 9 q^2}}{8} + (1 - x)\frac{
   p - q + \sqrt{-32 m^2 + 9 p^2 + 6 p q + 9 q^2} }{8}= \frac{q-p}{4}  \  \end{equation}
is always satisfying to $0\le x\le 1$, for $q\ge2m, \, p\ge2m$ and $m\ge0$. 

We close this section with an argument showing that the issue of an a priori unphysical stabilizer is
generic for any solvable system based on $\cV$ lying in a parabolic subgroup
$L_{\bf H} \ltimes N_{\bf H}\subset G_3$. In general, such a system is based on a graded decomposition 
\begin{equation}
\mathfrak{g}_3\cong  \overline{\mathfrak{n}}_{\bf H}
\oplus \scal{ \gl_1 \oplus \mathfrak{l}_{\bf H} \oplus \mathfrak{a}_{\bf H} } \oplus \mathfrak{n}_{\bf H} \ , 
\end{equation}
such that the $\gl_1$ generator ${\bf H}$ commutes with $\gl_1 \oplus \mathfrak{l}_{\bf H} \oplus \mathfrak{a}_{\bf H}$,
and all generators of $ \mathfrak{n}_{\bf H}$ are eigen vectors of strictly positive eigen value. For the system to admit
solutions parametrising the coset space $G_3/K^*_3$, ${\bf H}$ must moreover lie in $\mathfrak{k}^*_3$. As discussed in
this section, the asymptotic momentum if diagonalisable must necessarily lie in the $K^*_3$ orbit of the grade zero
component. Therefore its stabilizer in $K^*_3$ contains the $GL(1)$ subgroup generated by ${\bf H}$, and cannot be a
compact subgroup. For a total charge satisfying to \eqref{Kerr-Orbit}, the stabilizer in $Spin^*(16)$ will be either
$SU^*(8),\, SU(2,6)$ or $SU(4,4)$. The two last correspond to solutions with $\lozenge>0$ that violate the BPS bound.

\subsection{Extremal limits}
\label{sec:extremal}
This system interpolates between all the known solvable systems associated to nilpotent orbits of class $D_4$ \cite{Bossard:2012ge}. Indeed, 
the three-dimensional base metric is flat if one considers an extremal Euclidean electro-vacuum. Therefore, each of the
five classes of extremal Maxwell--Einstein instantons discussed below \eqref{eq:R-base} give rise to a solvable system
for extremal solutions. We summarise these five possibilities in table \ref{tbl:extremal}, giving the functions that
remain nontrivial in each case and the corresponding extremal systems they lead to.

%
\begin{table}[t]
\centering
{\renewcommand{\arraystretch}{1.7}
\begin{tabular}{|c|c|c|c|c|c|}
\hline
& BPS & \multicolumn{3}{c|}{Almost-BPS} & non-BPS 
\\ \hline
   & $\cE_+$, $K_+$ & $\cE_+$, $K_-$ \,& $\cE_-$, $K_+$ & $K_+$, $K_-$ & $\cE_-$, $K_-$ \\ \hline
Riemann & + & + & $-$ & neither & $-$
\\ \hline
Maxwell   & + & $-$ & + & neither & $-$ 
\\ \hline
\end{tabular}
}
\caption{A summary of the five possible extremal limits of the system. In each case we display
the functions that remain nontrivial and the selfduality of both the Riemann tensor and the Maxwell field strength of
the underlying Euclidean Maxwell--Einstein solutions.}
\label{tbl:extremal}
\end{table}

By construction, the system reduces to the almost-BPS system \cite{Goldstein:2008fq,Bena:2009ev,Bena:2009en} if one considers 
\begin{equation}\cE_+ = 1 \ , \qquad \cE_- = V^{-1} -1 \ , \quad K_- = 0 \ , \quad K_+ = K \ , \end{equation}
so that \eqref{eq:EK-eoms} imply that $V$ and $K$ are harmonic functions, which ultimately appear in the almost BPS
system. This is rather natural because considering $\cE_+$ to be constant with a nontrivial $\cE_-$ corresponds to
consider an anti-selfdual Gibbons--Hawking instanton. 

Accordingly, one gets back the BPS system \cite{Behrndt:1997ny} for 
\begin{equation}\cE_+ = V^{-1}  \ , \qquad \cE_- = 0 \ , \quad K_- = 0 \ , \quad K_+ = K \ , \end{equation}
where $V$ and $V K$ are again harmonic and correspond to turning on a selfdual flux in a selfdual Gibbons--Hawking space.
The various functions of the system are determined in terms of harmonic functions $\cH^\Lambda,\, \cK_\Lambda$ as
\begin{equation}\begin{split} 
V &= -\cH^0\,, \\
K &= - \frac{ \cH^1}{\cH^0}\,, \\
K^a &= - \frac{\cH^a }{ \cH^0} \,,
\end{split}\qquad\begin{split}
M &= \frac{1}{2} \scal{ \cH^0 \cK_0 + \cH^1 \cK_1 + \cH^a \cK_a} - \frac{ \cH^1 \cH^a \cH_a}{2 \cH^0} \,, \\
L &=   \cK_1 - \frac{ \cH^a \cH_a}{2 \cH^0}\,, \\
L_a &=   \cK_a - \frac{ \cH^1 \cH_a}{ \cH^0}\,. \end{split}
\end{equation}
We can also consider the case 
\begin{equation}\cE_+ = 1 \ , \qquad \cE_- = L_1^{\; -1}-1 \ , \quad K_- = -L_1^{\; -1}\, \bar K_1   \ , \quad K_+ = 0 \ , \end{equation}
in which case we recover the composite non-BPS system in a non-standard duality frame. The system takes the standard
form of \cite{Bossard:2011kz,Bossard:2012ge} in terms of the functions
\begin{equation}\begin{split} 
K^a &= \bar K^a - \frac{ L^a K_1}{L_1}\,, \\
L^a &= L^a\,, \end{split}\qquad\begin{split}
M &= \bar M - \frac{1}{2} \bar K_1 L_a L^a \,,  \\
L &=  \bar V - 2 \frac{\bar M \bar K_1}{L_1} + \frac{ \bar K_1^2\, L_a L^a}{2 L_1} \,.  \end{split}\end{equation}
The scalar fields are identical to the ones of the composite non-BPS system in the standard frame, up to the 
substitution 
\begin{equation}t^1 \rightarrow \frac{1}{\bar t^1}\ , \end{equation}
of the scalar parametrising the $SL(2)/SO(2)$ factor. This transformation is not inside $SL(2)$, and is in fact the 
combination of a Mobius inversion and the substitution $t^1 \rightarrow - \bar t^1$ that relates the BPS solutions to 
the non-supersymmetric extremal solutions for which the central charge vanishes at the horizon. So strictly speaking it 
is not the same composite non-BPS system in $\cN=2$ supergravity, but it would be the same up to an $E_{7(7)}$ duality 
transformation in $\cN=8$ supergravity.

The two remaining extremal limits both reduce to the almost BPS system in a non-standard duality frame. We shall only 
discuss the case of an Israel--Wilson base, which has been considered in detail in \cite{Bena:2009fi}. In this case one has
\begin{equation}\begin{split} 
\cE_+ &= \frac{m_-}{e_-} \frac{2}{K^\prime}-1 \,, \\
K_+ &=  \frac{m_-}{e_-} -\Scal{  \frac{m_-}{e_-}}^2 \frac{1}{K^\prime}\,,\\
K^a &= K^{\prime a} - \frac{L^a}{V^\prime K^\prime}\,, \\
M &= M^\prime  - \frac{1}{2} \frac{ L^{\prime a} L^\prime_a}{K^\prime}\,,
\end{split}\qquad\begin{split} 
\cE_- &= \frac{1}{V^\prime} - 1\,,\\
K_- &= \frac{e_-}{m_-} \Scal{ 2 - \frac{1}{V^\prime}}\,, \\
L_a &= \frac{m_-}{e_-} \frac{L^\prime_a}{K^\prime}\,, \\
L &= \frac{e_-}{m_-} \frac{ \frac{1}{2} L^{\prime a} L^\prime_a - 2 K^\prime M^\prime + K^{\prime\, 2} L^\prime 
V^\prime}{ V^\prime K^\prime}\,,
\end{split}\end{equation}
where the prime functions satisfy to the almost BPS equations. The embedding of the system is related to the standard 
one by an $SL(2)$ duality transformation acting on the $t^1$ modulus as
\begin{equation}t^{\prime 1}  = -  \Scal{  \frac{m_-}{e_-}}^2 \frac{1}{t^1} + \frac{m_-}{e_-} \ , \end{equation}
which is the combination of a Mobius inversion, an axionic shift, and a rescaling.

\subsection{Five-dimensional uplift}
\label{sec:5D-lift}
We now consider the uplift of the above system to five dimensions. In this case the metric takes the 
form
\begin{equation}\label{eq:5D-metr}
 ds^2_5 = -\scal{\tfrac{1}{2} L L_a L^a}^{-\frac{2}{3}} \scal{dt + k}^2 +\scal{\tfrac{1}{2} L L_a L^a}^{\frac{1}{3}}  
\scal{ {V}^{-1}\, \scal{ d\psi + w^0 }^2 + V\, \gamma_{ij}dx^idx^j } \,,
\end{equation}
which is the standard timelike fibration in the floating brane Ansatz, over the four-dimensional Euclidean base 
solution of Euclidean Maxwell--Einstein equations that has
now become physical. The one-form $k$ is given by
\begin{equation}\label{eq:5d-k-gen}
 k = \omega -\frac{M}{V}\, ( d\psi + w^0 ) \,.
\end{equation}
The gauge fields read
\begin{align}\label{eq:5d-gauge-gen}
A^1 =&\, \frac{1}{L} \scal{ dt + k} +  w^1 + K_+\,\scal{ d\psi + w^0 }  \,,
\CR
A^a =&\, 2 \frac{L^a}{L^b L_b}\scal{ dt + k} +  w^a + K^a\,\scal{ d\psi + w^0 } \,,
\end{align}
while the scalar fields read
\begin{equation}\label{eq:5D-scal}
X^1 = L^{-\frac{2}{3}} \scal{ \tfrac{1}{2} L_a L^a}^\frac{1}{3} \, ,
\qquad
X^a = L^\frac{1}{3}   \scal{ \tfrac{1}{2} L_b L^b}^{-\frac{2}{3}} L^a \,, 
\end{equation}
so that $\tfrac{1}{2} X^1 X_a X^a =1$, as is required.

In five spacetime dimensions, it is natural to define the projection of the
gauge field strengths on the Euclidean base, given by
\begin{align}
F^1=d\left( \frac{1}{L}(dt + k) \right) + \Theta^1\,,
\quad \Rightarrow\quad
\Theta^1=d\scal{ w^1 + K_+ \,( d\psi + w^0 )}\,,
\CR
F^a=d\left( \frac{L^a}{\tfrac12\,L^b L_b}(dt + k) \right) + \Theta^a\,,
\quad \Rightarrow\quad
\Theta^a=d\scal{ w^a + K^a \,( d\psi + w^0 )}\,.
\end{align}
The Euclidean field strengths $\Theta^1$, $\Theta^a$ are then given by 
\begin{eqnarray}\label{eq:5d-flux}
\Theta^1 &=&d K_+\wedge( d\psi + w^0 )  +  V\, \star  d K_+ \ ,
 \CR
\Theta^a 
&=& d K^a \wedge( d\psi + w^0 )    + V\, \star \scal{ d K^a -2\, L^a d K_-}\,,
\end{eqnarray}
where $\Theta^1$ is manifestly selfdual, whereas the anti-selfdual components of the $\Theta^a$ arise
due to the non-trivial function $K_-$, that determines the anti-selfdual component of the Euclidean
electro-vacuum.

With these definitions, we compute that once a Maxwell--Einstein solution is
specified by choosing the functions $\cE_\pm$ and $K_\pm$, the remaining
functions $K^a$, $L_a$, $L$ are found by solving the following equations
\begin{eqnarray}
\nabla^2 K^a &=& 2\,\nabla L^a \cdot \nabla K_- + 2\,V\,(\nabla \cE_+ + K_- \nabla K_+)\cdot \nabla K^a
                 -2\,V^{-1}\,  L^a\,  (\nabla \times {w}^0) \cdot \nabla K_- \CR
\nabla^2 L_a &=& 2\, V\, \nabla K_+ \cdot \left( \nabla K_a - L_a\,\nabla K_- \right) \CR
\nabla^2 L &=&  
 V\, \nabla K^a  \cdot (  \nabla K_a - 2 \nabla L_a  )
    -2\,\upkappa  \cdot \nabla K_-  \,,
\end{eqnarray}
where the vector, $\upkappa$, determines the anti-selfdual component of the vector $k$ as
\begin{eqnarray}
dk-\star_4dk &\equiv& \upkappa \wedge ( d\psi + w^0 )  - V \star \upkappa \CR
&=&  d\omega + \star d M + 2\,V\, M \, (\star d\cE_- + K_+ \star d K_-)  -\text{dual}
\end{eqnarray}
whereas its selfdual component reads
\begin{equation}dk+\star_4 dk = - V\, \scal{ L \, \star d K_+ + L_a \, d K^a  - L_a L^a \,\star d K_- } + \text{dual} \ . \end{equation}
It is straightforward to verify that this system, as written above in five-dimensional supergravity, is
equivalent to the system obtained in \cite{Bena:2009fi} using the floating brane Ansatz.

\section{The bubbling bolt solution}\label{sec:2cen-ex}

We now turn to an explicit example of a solution to the system of the previous section, using the
results of section \ref{sec:instanton} on multi-centre Maxwell--Einstein instantons. The simplest
case is that of a two-centre solution, where one only considers the non-extremal centre and a
single extremal centre. As it turns out, it is not possible to support proper horizons on this
base, but it is possible to obtain a singular four-dimensional solution that lifts to a smooth
five-dimensional geometry, in exactly the same way as for multi-centre BPS composites 
\cite{ Bena:2005va,Berglund:2005vb}. This explicit construction allows to show that all essential
properties of extremal microstate geometries carry over to the non-extremal case, despite the caveat of
unphysical asymptotics.
We present the solution and the method of obtaining it in subsection \ref{sec:sol}, while in
subsection \ref{sec:5d-sol} we consider the five-dimensional uplift of the solution, in order to
show that it is everywhere smooth and free of closed time-like curves in that setting. We investigate
the BPS bound for the smooth geometries obtained and show that it is always broken. Finally, in
section \ref{sec:fluxes} we discuss the fluxes on the two-cycles of the solution, for completeness.

\subsection{The solution}\label{sec:sol}
Despite the fact that the system is solvable by construction, it is rather involved in practice
to obtain an explicit solution by integrating the equations directly. This is partly due to the
fact that there exist harmonic functions on the base under consideration, that are however not
rational functions of the distances from the centres. We do not allow for solutions exhibiting
such behaviour, both for simplicity and to avoid singularities at the centres.

The method we use to solve the system for the case of two centres is based on expanding the
scalar momentum on a basis of conserved vector fields, which are all based on rational
functions, as
\begin{equation}\label{eq:mom-expand}
\cV P \cV^{-1} =  \mathcal{J}_0  \frac{dr}{r^2-c^2} +  \mathcal{J}_\pA \frac{ \scal{ r + \frac{c^2}{m_-}}^2}{r^2-c^2} 
d\cH_\pA + \mathcal{J}_{0,\pA} \Scal{ \frac{   r + \frac{c^2}{m_-}}{r^2-c^2} d\cH_\pA + \cH_\pA   \frac{dr}{r^2-c^2}} 
\,.
\end{equation}
Here, $\mathcal{J}_0$, $\mathcal{J}_\pA$ and $\mathcal{J}_{0,\pA}$ are constant vectors taking
values in the Lie algebra $\mathfrak{g}_3\cong\so(4,2+n)$, while the $\cH_\pA$ are the functions \eqref{eq:H1},
describing the various extremal centres, in the general case. Each of the three conserved vector
fields in \eqref{eq:mom-expand} has a clear physical meaning, since the first two terms are the
vector fields arising from a pointlike source at each of the non-extremal and extremal centres,
while the third term is a dipole between the non-extremal and extremal centres. Therefore, there
is no interaction between extremal centres within this Ansatz, which is sufficient for the present
example, where we consider a single extremal centre. 

In practice, the above strategy consists in imposing an expansion as in \eqref{eq:mom-expand} for each of the
vector fields in \eqref{eq:alm-NE-mag}-\eqref{eq:alm-NE-el} and solve the resulting equations as
a linear algebraic system for the functions $K^a$, $L_a$, $L$ and $M$ and their derivatives. The
functions $\cE_\pm$, $K_\pm$, $V$ and the vector field $w^0$ are taken to be the ones in
\eqref{eq:SL3-fun-mult}, \eqref{eq:V-mult} and \eqref{eq:w0gen} respectively, where we restrict ourselves
to a single extremal centre at a distance $R>c$ from the extremal centre. Starting with the functions
$L_a$, $K^a$, we obtain the result
\begin{align}\label{LKsol} 
L_a=&\,\frac{(m_{-}+r) \left(c^2+m_{-} r\right)}{2 m_{-} \left(r^2-c^2\right)}\frac{l_a}{V} + u_a \,,
 \nn\\
K^a =&\, \frac{2 p^a}{m_{-}-m_{+}}
-\frac{2 e_{-} \left(c^2+m_{+} r\right) u^a}{(m_{-}-m_{+})\left(c^2+m_{-} r\right)}
-\frac{2 e_{-} \left(r^2-c^2\right)}{(m_{-}+r) \left(c^2+m_{-} r\right)}\,L^a \,,
\end{align}
where we use the metric \eqref{eq:STU-eta} to raise and lower indices. In these expressions, the $p^a$, $u_a$,
are constants parametrising the charges and $l_a$ are additional integration constants.
The latter arise due to the fact that some of the equations in the linear algebraic system we consider may
be dependent, so that they must be integrated explicitly as differential equations, and this is exactly what
happens in the case above. Luckily, the system involving the more complicated functions $L$ and $M$ does
not suffer from this complication and one straightforwardly obtains the following result by algebraic manipulation 
\begin{align}
L =&\,\frac{4 e_- (c^2 - r^2)}{(m_- + r) (c^2 + m_- r)} M 
     \nn\\
 &\,
 - \frac{2 e_-^2 (c^2 - r^2)^2}{(m_- + r)^2 (c^2 + m_- r)^2}\,V\,
   \left( L_a L^a + \frac{(m_- + r)  (c^2 - r R)}{(r^2 - c^2) (m_- + R)}\,u_a u^a \right)\,,
\\
M=&\,
-\frac{e_- (r^2-c^2 ) }{(m_- + r) (c^2 + m_- r)}
    V \, L_a \left(L^a  -u^a\right) 
    \nn\\
 &\,
    +\frac{e_-}{2\,(m_- + R)} \,
    \left( \frac{R - r}{m_- + r}\, V 
          +\frac{c^2 +  m_- r}{ 2\,c^2 (c+m_-)^2 (r^2 - c^2)} \,(f_1 r + f_2) \right)\, u_a u^a\,,\label{Msol}
\end{align}
where the constants $f_1$, $f_2$ appearing in $M$ are such that
\begin{align}
f_1 r + f_2= &\,\left( c^2 - m_- m_+ -\frac{2\,c^2\,k\,n_1}{m_- + R} \right)\,
     \left( (R - c)^2 + R (r - R) \right) 
     \nn\\
 &\,
 + \left( c^2 (m_{+} + m_{-} ) + 2 c m_{-} m_{+} - \frac{2\,c^2 m_-\,k\,n_1}{m_- + R}\right)(r - R) \,.
\end{align}
Finally, \eqref{eq:do} gives the angular momentum one-form as
\begin{equation}
\omega= -\frac{e_{-} R\, u_a u^a}{2\,(R+m_-)^2}\, \left( \Scal{1 - \frac{r + R }{r_1}}\,(1 - \cos{\theta}) 
  +  \frac{c^2}{R\,r_1} \sin^2{\theta} \right) d\varphi\, .
\end{equation}
Here we enforced in the Ansatz that $\omega$ was globally defined and therefore free of Dirac string singularity. In the asymptotic region we get 
\begin{equation}
\omega= \frac{e_{-} (R^2-c^2) \, u_a u^a}{2\,(R+m_-)^2}\,  \frac{\sin^2{\theta}}{r} d\varphi + \mathcal{O}(r^{-2}) \, ,
\end{equation}
such that the angular momentum of the solution is 
\begin{equation}\label{eq:Jval} J = \frac{e_{-} (R^2-c^2) \, u_a u^a}{2\,(R+m_-)^2}\ . \end{equation}
With these data, one may now construct the metric through the standard Ansatz
\eqref{eq:eU}-\eqref{eq:do}.

The above solution of four-dimensional supergravity describes a system of two centres without horizon. The scale factor diverges as $e^{-4U}\sim 1/r$ near each centre, so that the solution
is strictly speaking singular from a four-dimensional point of view. However, we intent to
interpret this solution as a microstate geometry associated to a four-dimensional black hole, which is smooth and free of closed time-like curves as a five-dimensional geometry, as explained in the next subsection. With this goal in mind, we are only interested in the total electromagnetic charges of the solution in four dimensions. The magnetic charges read
\begin{align}\label{eq:4d-ch-m}
P^0 = &\, (n + n_1)\, k\,,
\CR
P^1 = &\, -\frac{c^2 - m_-^2}{2\, e_-} - \frac{m_-}{e_-}\, P^0  \,,
\CR
P^a = &\, \frac{2}{m_+ -m_-}\,\left(p^a - e_- u^a \right)\,P^0  - e_- u^a\,,
\end{align}
while the electric charges take the form
\begin{align}\label{eq:4d-ch-e}
Q_1 = &\, \frac{1}{2\,P^0}\, P^a P_a
\CR
 &\,+ \frac{2\,e_-^2\,u^a u_a}{(c + m_-)^2}\, \frac{1}{P^0}\left( 
     \frac{(R-c)^2}{R\,(R+m_-)}\,\scal{P^0 + \tfrac12\, (m_+ - m_-)}P^0 -\left(P^0 - \tfrac12\, (c + m_-)\right)^2 \right) 
\,,
\CR
Q_a = &\, \frac{1}{P^0}\, \scal{ P_a + e_- u_a}\, P^1 = \frac{2}{m_+ -m_-}\,\scal{ p_a - e_- u_a }\, P^1\,,
\CR
Q_0 = &\,-\frac{1}{2\,(P^0)^2} \Scal{ P^a + e_- u^a }\Scal{ P_a + e_- u_a }\, P^1
\CR
 &\,+ \frac{2\,e_-\,m_-\,u^a u_a}{(c + m_-)^2}\, \left( 
     \frac{(R-c)^2}{R\,(R+m_-)}\,(P^0 + \tfrac12\, (m_+ - m_-) ) -\left(P^0 +\frac{c^2 - m_-^2}{2\, m_-}\right) \right) 
\,.
\end{align}
We refrain from giving the individual charges at the two centres, but we stress that $n_1$ is the only
independent charge at the extremal centre, whereas all other charges are completely fixed in terms of the charges
at the non-extremal centre. The quartic invariant of the total charges reduces to 
\begin{multline} \label{eq:I4-sol}
I_4(Q,P) = - \frac{ e_-^{\; 2} k\, n_1 (R-c)^2 \, (u_a u^a)^2}{(c+m_-)^3(R+m_-)^4} 
\Bigl( k\, n_1 (R-c) (c+m_-)^2 \scal{ (c+m_-) (R-c) + 4\, c\, m_- } \\+ 2\, c^2 (R+m_-)^2 \scal{ (c+m_-)^2 - 4\, k\, m_-} \Bigr)  \ .
\end{multline}
Note that the quartic invariant does not depend on the parameters $p^a$, which only appear in the definition of the charges as a T-duality parameter. As explained in section \ref{sec:BPSbound}, the system at hand only has solutions with total electromagnetic charges admitting a strictly negative quartic invariant. This is not yet manifest in \eqref{eq:I4-sol}, but we shall see in the next section that the positivity of the dilaton fields requires through \eqref{eq:reg-conds} that $n_1$ and $n$ are positive integers and
\begin{equation}(c+m_- )^2 > 4 \, k \, m_-  > 0 \ , \end{equation}
such that $I_4(Q,P) $ is indeed strictly negative. 

Because $n_1> 0$, one finds that $V$ in \eqref{eq:V-mult} vanishes on a surface
in the three-dimensional base. As explained in more detail in the next subsection, where we consider
the five-dimensional uplift, the solution is smooth on this surface, which is an evanescent ergo-surface.
In contrast, there is no special physical feature of the four-dimensional solution at that surface, as
we have checked in numerical examples.

A final remark is in order, since the solution above is not the most general obtained by the procedure
outlined in the beginning of the section. In particular, there are more general solutions where the
constant, $J$, is not restricted as in \eqref{eq:Jval}, so that it represents an additional parameter
that appears in $L$ and $M$ nontrivially. In these solutions, the scale factor diverges as
$e^{-4U}\sim 1/r^2$ near the extremal centre, corresponding to a two-charge, rather than a one-charge
centre. We have been unable to find an example without closed timelike curves in this more general
class and we believe it to be unlikely.

\subsection{Smoothness and closed time-like curves}\label{sec:5d-sol}
We shall now study the five-dimensional uplift of the solution, and constrain the parameters to get a smooth geometry 
free of closed time-like curves. The five-dimensional fields are defined as in section  \ref{sec:5D-lift}, and the 
metric is in particular defined from \eqref{eq:5D-metr} with the four-dimensional Euclidean metric defined in section 
\ref{sec:instanton}. Henceforth, we set $n_1=1$ for simplicity, so that there is no orbifold singularity at the
extremal centre and the geometry near $r=R$ is that of flat space. Note that there is no regular solution without 
closed time-like curves for $n_1=-1$. We will also set $k=1$, without lost of generality, since this parameter only 
parametrises a scale associated to the coordinate $\psi$. 

Regularity at each centre requires that $M$ has no pole at $r=c$ and $r_1=0$. In fact we have already constrained the 
solution \eqref{Msol} to satisfy to this criterion. Moreover, $L,\, L_a$ are finite at the poles, as well as the vector 
$\omega$. Therefore the singularities of the metric at the poles are determined by the ones of the Euclidean 
four-dimensional metric \eqref{eq:base-metr} that we have already discussed in section \ref{sec:instanton}. We therefore 
consider $m_+$ as determined by \eqref{mm}, and use \eqref{eq:Dir-bolt} to determine 
\begin{equation}R = - m_- + \frac{2 (c + m_-)^2 }{(c+m_-)^2 - 4 c  - 2  (c + m_-) n} \  . \end{equation}
The condition $R>c$ then already gives a non-trivial bound on $m_-$. 

In order to have a physical solution, one must moreover impose that no closed time-like curves appear. We already 
enforced $\omega$ to be globally defined through \eqref{eq:Jval}, and we further require that 
\begin{equation}e^{-4U} -\frac{ \omega_\varphi^{\; 2} }{(r^2-c^2) \sin^2\theta} > 0 \ . \label{OmegaStrong} \end{equation}
to ensure that the orbits generated by $\partial_\phi$  are space-like. In order to check that the orbits generated by 
the Killing vector $\partial_\psi$ are also  space-like, it is convenient to rewrite the metric \eqref{eq:5D-metr} as a
$U(1)$-principal bundle with respect to this isometry, \ie 
\begin{equation}
 ds^2_5 = e^{2U} V \scal{\tfrac{1}{2} L L_a L^a}^{\frac{1}{3}}   \, ds^2_4 + e^{-4U} V^{-2} \scal{\tfrac{1}{2} L L_a 
L^a}^{-\frac{2}{3}}   \, \scal{d\psi + A^0 }^2\,,
\end{equation}
where the four-dimensional metric is \eqref{FourDmetric} and $A^0$ is the four-dimensional vector field 
\begin{equation}A^0 = e^{4U} VM ( dt + \omega ) + w^0 \ . \end{equation}
The absence of time-like closed curves along this fibre is ensured by the conditions \cite{Bena:2005va}
\begin{equation}e^{-4U} > 0 \ , \qquad V L > 0 \ , \qquad V L_a > 0 \ . \label{CTCs} \end{equation}
Enforcing \eqref{CTCs} at the poles, one obtains the bounds 
\begin{eqnarray}\label{eq:reg-conds}
&& 1+n<c \le 2+n \ , \qquad n + \sqrt{ 4 c +n^2}-c< m_- < c \ , \qquad l_a>0 \,,
\CR
&& -\frac{c\scal{ c^3 + 2 c (c+n)  m_- +(c+2n-4)   m_-^{\; 2} }}{(c+m_-^{\; 2})^2} l_a  < u_a<0 \,.
\end{eqnarray}
and the condition that $n\in \mathds{N}$. Provided these conditions are satisfied, $V L,\, V L_a$ are in fact positive 
everywhere. 

Because $n_1=1$ is positive, $V$ admits a zero locus and the four-dimensional Euclidean base metric \eqref{eq:5D-metr} 
changes its overall sign as one crosses the surface  $V=0$, so that the base metric is  ambi-polar.  Therefore we must
also analyse the properties of the solution at the dangerous locus $V=0$, which is given in general by
\begin{equation}\label{eq:v0}
\frac{4\,k^2n_1^2}{r_1^2}\,(R^2 - c^2)=\frac{(r+m_+ )^2(R+m_- )^2 - 4\,c^2 k^2 n_1^2}{r^2 - c^2}\,.
\end{equation}
At fixed time $t=t_0$, this describes a three-dimensional surface in the four-dimensional  space-like section 
$t=t_0$, which is necessarily compact, since there is no solution of \eqref{eq:v0} for $r_1\sim r\rightarrow \infty$.
In particular, this surface necessarily encloses the extremal centre and may or may not in principle enclose the
non-extremal centre, depending on the parameters. In practice \eqref{eq:reg-conds} already implies that this surface
only encloses the extremal centre. Note that \eqref{OmegaStrong}
and \eqref{CTCs} also ensure that the differential $dt$ is still time-like at the evanescent ergo-surface,
such that the constant time $t=t_0$ slices are everywhere space-like Cauchy surfaces. 
   
Despite the fact that this is a non-extremal solution, the mechanism discussed in \cite{Bena:2005va,Berglund:2005vb} is 
still at work in exactly the same way, leading to a regular solution. Indeed, it is simple to verify that the $V L_a$
and $V K^a$ are linear functions of $V$, while $V\, L$ and $V M$ are quadratic functions in $V$, which are positive 
definite provided  \eqref{eq:reg-conds} is satisfied. The various components of the metric and the gauge fields are
such that they remain regular at the locus $V=0$, in the following way. Starting with the metric components along
$d\psi$, one can show they are regular on this surface, since
\begin{equation}
g_{t\,\psi} = -(\tfrac12\,L L^aL_a)^{-2/3}\,\frac{M}{V}= \frac{V\,M}{(\tfrac12\,V^3 L L^aL_a)^{2/3}}\,,
\qquad
g_{\psi\,\psi} = \mathcal{O}(V^0)\,,
\end{equation}
\ie $g_{t\,\psi}$ is everywhere regular by construction and $g_{\psi\,\psi}$ has no pole near the
$V=0$ locus. Similarly, the poles of the gauge fields at $V=0$ cancel identically as
\begin{equation}
A^1 \sim \frac{M}{V\,L} - K_+ = \mathcal{O}(V^0) \,,
\qquad
A^a \sim \frac{M\,L^a}{V\, (\tfrac12\,L^bL_b)} - K^a = \mathcal{O}(V^0) \,,
\end{equation}
near that surface. Finally, the scalar fields \eqref{eq:5D-scal} are invariant under a rescaling
of the $L^a$ and $L$, so that the pole at $V=0$ cancels identically. 

At this point we still need to check that  \eqref{OmegaStrong} is true everywhere, and in particular at the poles, the ergo-surface and in the asymptotic region. However this condition turns out to be rather difficult to solve analytically. One can in principle solve analytically for the condition $e^{-4U}>0$ at both the ergo-surface and the asymptotic region, but the explicit bounds on $c,\, m_-$ and $u_a$ turn out to be rather complicated and not very suggestive. These bounds are strictly stronger than \eqref{eq:reg-conds}, and one finds a narrow window of allowed parameters around the specific solution
\begin{equation}c = \frac{3}{2} + n \ , \qquad m_- = 1+n \ ,\qquad  u_a = - \scal{ \tfrac{4}{3} n^2 + 2 n } l_a \ , \label{RegSol} \end{equation}
that indeed satisfies $e^{-4U}>0$ everywhere for all $n\in \mathds{N}^*$. Having in mind the interpretation of the solution has representing a four-dimensional black hole microstate, we fix $U$ to zero at infinity, which determines $l_a l^a\, e_-$ as a function of 
$n$ only. This function is not particularly illuminating, and we only display its leading expansion at large $n$ 
\begin{equation} l_a l^a \, e_- = \frac{36 \, n}{\sqrt{5}} + \mathcal{O}(n^0) \ . \end{equation}
For the example solution \eqref{RegSol}, the complicated function \eqref{OmegaStrong} is parametrised by $n$ only. Although we did not manage to prove that  \eqref{OmegaStrong} is then identically satisfied analytically, we checked it numerically for a large sample of values of $n$ from $1$ to $1000$. In all specific examples we checked explicitly we find that
the contribution from $\omega_\varphi^{\; 2}$ is very small, and it would be very unlikely that some solutions in
this class do carry such closed time-like curves at unexpected loci.

Since we would like these smooth solutions to describe four-dimensional black holes, it is natural to look at the large 
$n$ limit in which the black hole is macroscopic, while the compactification radius modulus remains finite as
\begin{equation}\lim_{r\rightarrow \infty}  e^{-2U} V^{-1} \scal{\tfrac{1}{2} L L_a 
L^a}^{-\frac{1}{3}} = \sqrt[3]{\frac{2}{3}} + \mathcal{O}(n^{-1}) \ . \end{equation}
However, one would expect then the two centres to be localised whereas one gets 
\begin{equation}R-c = 6 n + \mathcal{O}(n^0) \ , \end{equation}
which exhibits that this solution deviates from a potential black hole solution at a large distance from the bolt.

One finds that the solution is still rotating asymptotically along the fibre, with
\begin{equation}k_\psi=- M/V= 2\sqrt{5}  + \mathcal{O}(n^{-1}) \ , \end{equation}
but with a negligible momentum in the large $n$ limit.  Whereas the angular momentum in four dimensions scales as
\begin{equation}J = \frac{6 n}{\sqrt{5}} + \mathcal{O}(n^0)\ . \end{equation}
Considering $p^a = \mathcal{O}(n^0)$, the only charges of order one in $n$ are 
\begin{equation}Q_a = \frac{2n}{3} l_a  + \mathcal{O}(n^0) \ , \qquad P^0 = n+1  \ , \qquad P^1 = - \frac{n\sqrt{5}}{36} l_a l^a  + 
\mathcal{O}(n^0) \ .  \end{equation}
Note that the solution is regular independently of the explicit values of the parameters $l_a$ and $p^a$, provided 
$u_a$ and $e_-$ are determined in terms of $l_a$ as above. These parameters then only enter the solution through
$SO(2,n)$  duality transformations, and one checks indeed that $I_4(Q,P)$ is a function of $n$ only  
\begin{equation}I_4(Q,P) = - \frac{144\, n^3 }{5} + \mathcal{O}(n^2)  \ . \end{equation}

The final test for the admissibility of this solution as a microstate geometry is the BPS bound on the mass.  Its 
verification is a complicated task in the general case, but it is simple to analyse for
any particular example. In the large $n$ limit described above we find that the central
charges and the four-dimensional ADM mass exhibit the following hierarchy \footnote{With in general $2 |Z_\deux|^2= Z_a 
\bar Z^a + \sqrt{ ( Z_a \bar Z^a)^2 - |Z_a Z^a|^2}$.}
\begin{equation}
 11n > |Z_\un|> M_{\rm \scriptscriptstyle ADM} >  |Z| > 8n > |Z_\deux|=|Z_\trois|  \,,
\end{equation}
so that the BPS bound is violated in $\cN=8$ supergravity, although the $\cN=2$ supergravity BPS bound is strictly speaking still satisfied. We checked numerically that the same ordering holds for all $n\in \mathds{N}^*$. Since $I_4(Q,P)$ is only of order $\mathcal{O}(n^3)$ 
\begin{equation}W = |Z_\un| + \mathcal{O}(n^0)\ , \end{equation}
and because $I_4(Q,P)<0$ for all $n$, $M_{\rm \scriptscriptstyle ADM}<|Z_\un|< W$ consistently with our general analysis in section \ref{sec:BPSbound}. Moreover, we checked explicitly the characteristic equation \eqref{Kerr-Orbit} for all $n$, which implies that all asymptotic scalar momenta $dt^1, \, dt^a$ are determined by the asymptotic charges and the ADM mass, just as for a regular black hole solution (modulo the specific branch as discussed in section 3.2).

A pure state in a conformal field theory certainly does not violate the BPS bound, 
and therefore this solution should not be admissible as an actual black hole microstate. A violation of the BPS bound 
was already pointed out in a smooth solution carrying a bolt 
\cite{Compere:2009iy}. It was understood in \cite{Gibbons:2013tqa} that this bolt defined a self-intersecting homology 
sphere, and therefore an obstruction to the existence of spin structure on the manifold. In our case we have also seen
in section \ref{sec:spin} that the considered space-time does not admit a spin structure, although for even $n$ it is
the homology sphere linking the nut to the bolt that leads to the obstruction.

\subsection{Description of the fluxes} \label{sec:fluxes}
For completeness, we also discuss the structure of the gauge field fluxes in the five-dimensional
uplift of the solution. The base space field strengths, $\Theta^1$, $\Theta^a$, are defined from \eqref{eq:5d-flux}, but it is convenient to use the explicit solution \eqref{LKsol} to rewrite $\Theta^a$ as 
\begin{eqnarray} \label{eq:flux-expl}
\Theta^a &=& u^a\, \scal{ dK_-\,(d\psi + w^0) - V\, \star d K_-} + 2 \, \frac{ p^a-e_- u^a}{m_+-m_-} dw^0 \CR
&& \hspace{35mm}  - d \Bigl[  \Scal{ 2 \, \frac{ p^a-e_- u^a}{m_+-m_-} + \frac{e_-}{m_-} l^a V^{-1} } ( d\psi + w^0) \Bigr] \ , 
\end{eqnarray}
We observe that the constants $l^a$ only appear through the total derivative of a globally defined
one-form so that they do not appear in any fluxes, consistently with the fact that they also
do not appear in the charges \eqref{eq:4d-ch-m}-\eqref{eq:4d-ch-e}, as computed in four dimensions. 

Indeed, the charges of the solution are supported by the two different two-cycles present in the geometry.
First, the non-extremal centre at $r=c$ is a $S^2$ bolt carrying a NUT charge $n$. The fluxes of
the magnetic field strengths $\Theta$ over the bolt are given by
\begin{align}\label{eq:bolt-flux}
\frac1{4\pi}\int_{\text{B}}\!\Theta^a = &\,
-\frac{2\,k\, n}{m_{+}-m_{-}} p^a
-\frac{2\,e_{-} m_{+}\,k\,n_1 }{(m_{+}-m_{-})(R+m_{-})}\, u^a\, = p^a + \frac{3\, l^a}{\sqrt{5}\, l_b l^b}  + \mathcal{O}(n^{-1})   ,
\CR
\frac1{4\pi}\int_{\text{B}}\!\Theta^1 = &\,
\frac{m_-}{e_-}\,k\, n +\frac{c^2-m_{-}^{\; 2} }{2 e_-} = \frac{\sqrt{5}\, n}{36} \, l_a l^a + \mathcal{O}(n^{0})  \,,
\end{align}
where we also give their large $n$ expansion  associated to the explicit solution discussed in the last section. Note that because $p^a$ are arbitrary, the fluxes of $\Theta^a$ on the bolt are not constrained, whereas the fluxes of $\Theta^1$ on the bolt are determined to be a specific function of $n$ multiplying $l_a l^a$.

Similarly, one can consider the fluxes of the magnetic field strengths $\Theta$ over the nontrivial
two-cycle, $\Delta$, swept out by the $U(1)$ fibre generated by $\partial_\psi$ between the two points where it collapses, namely
the extremal centre and the (north pole of the) bolt at the non-extremal centre. The computation is
greatly simplified by noting that the vector fields, including $w^0$, can be made regular on the line
connecting these two points and therefore one can compute the flux by
\begin{equation}
\frac1{4\pi}\int_{\Delta}\!\Theta=
\frac1{4\pi}\left(\int_{0}^{4 \pi k}\!A \bigr|_{r=c}\,d\psi  -\int_{0}^{4 \pi k}\!A \bigr|_{r=R}\,d\psi \right)
= k\left( K \bigr|_{r=c} - K \bigr|_{r=R} \right) \,,
\end{equation}
from which we obtain the expressions
\begin{align}\label{eq:bubble-flux}
\frac1{4\pi}\int_{\Delta}\!\Theta^a = &\,
\frac{ 2 \,k\,e_{-} (R - c)}{(c+m_{-})(R+m_{-})}\,u^a = - \frac{18\, l^a}{\sqrt{5}\, l_b l^b}  + \mathcal{O}(n^{-1}) \,,
\CR
\frac1{4\pi}\int_{\Delta}\!\Theta^1 = &\, -\frac{(c+m_{-})^2}{4\,e_-} =  -\frac{\sqrt{5}\, n }{36} \, l_a l^a + \mathcal{O}(n^{0})\,.
\end{align}
Again these fluxes are determines as specific functions of $n$ multiplying respectively $\frac{l^a}{ l_b l^b}$ and $l_a l^a$. It seems that the repulsion between the bolt and the nut is mostly supported by the flux of $\Theta^1$ on the bubble joining them.  Note that the $p^a$ appear explicitly in \eqref{eq:bolt-flux}, despite the fact that they drop out
from the field strengths, due to the nontrivial flux of the $w^0$ gauge field on the bolt. This is
clear from \eqref{eq:flux-expl}, where we have rewritten the flux explicitly in terms of $w^0$ and
the relevant globally defined total derivative.

\section{Conclusion}
\label{sec:concl}
In this paper, we investigated a solvable system of equations, which solves the equations of motion of
five-dimensional $\cN\!=\!1$ supergravity. This system includes non-extremal solutions, as well as all 
the known solvable systems describing multi-centre extremal black hole solutions, in appropriate
limits. We have found that this kind of system necessarily leads to an asymptotic behaviour such that the asymptotic scalar momenta are always off the attractor flow of the regular non-extremal black hole solution carrying the same total charges and mass. We studied a particular example that reduces to the floating brane Ansatz in a Euclidean
electro-vacuum background \cite{Bena:2009fi}. 

We have shown that there exists a large class of ambi-polar Maxwell--Einstein instantons that include
a non-extremal bolt and an arbitrary number of Gibbons--Hawking centres. For each such instanton
solution, there are functions satisfying to a solvable system of differential equations, which define
a complete multi-centre supergravity solution. As an explicit example, we have presented a family of
two-centre smooth solutions, describing a non-extremal charged bolt interacting
with a Gibbons--Hawking centre. Although these two-centre solutions define smooth globally hyperbolic
space-times, their ADM mass turns out to violate the BPS bound. This situation is similar to the
violation of the BPS bound by smooth solutions derived in \cite{Compere:2009iy}, which was already
understood in \cite{Gibbons:2013tqa} to be a consequence of the absence of spin structure on the
corresponding space-time. We similarly demonstrate that the class of Maxwell--Einstein instantons we
use in order to define the Euclidean base space never admits a spin structure. It is therefore possible that generalisations of our two-centre solutions will eventually always violate the BPS bound.

Despite the negative results on the asymptotic properties of the given solution, there are several lessons to take away
from this exercise. Indeed, the results of this paper show that there exists a large variety of
non-extremal smooth globally hyperbolic space-times solutions that consist of a non-extremal bolt
fixed point and an arbitrary number of Gibbons--Hawking nut fixed points. These solutions would
appear as good candidates for microstate geometries associated to non-extremal black holes if they
were not violating the BPS bound. It seems necessary to explicitly impose that a microstate
geometry must admit a spin structure. This is a rather minimal exigence in supergravity to require
that all fundamental fields can be defined globally in a vacuum background, and this permits to
disregard all solutions violating the BPS bound, which would presumably be inconsistent as quantum states in
string theory. Note moreover that the five-dimensional theory is ungauged, and there is no composite
connection associated to the scalar fields parametrising $SO(1,1) \times SO(1,n)$ that could permit
to consider a spin$^{\rm c}$ structure. With this requirement, all the solutions based on the
Maxwell--Einstein instantons we have defined in this paper are ruled out, since the presence
of a bolt fixed point and a Gibbons--Hawking centre together is enough to forbid the existence of
spin structure. To ensure the existence of spin structure, it seems that one should instead
consider a Euclidean base space including both selfdual and anti-selfdual Gibbons--Hawking
centres.

More generally, it was shown in section \ref{sec:NE-sys} that solvable systems of the type
constructed in this paper may only contain solutions with an asymptotic structure that is
necessarily different form that of regular non-extremal black holes, irrespective of whether they
respect the BPS bound or not. Indeed, it turns out that, even for solutions that have the mass
and charges of a regular black hole, the scalar momenta at infinity cannot match the corresponding attractor flow.
 This is in fact a general property of  all known non-extremal
globally hyperbolic smooth solutions 
\cite{Jejjala:2005yu,Bena:2009qv,Bobev:2009kn,Compere:2009iy}. This may turn out to be a general general property of globally hyperbolic smooth geometries, and it would be interesting to investigate if this could be a consequence of the generalised Smarr formula \cite{Gibbons:2013tqa,Haas:2014spa}.

One may wonder about the possible interpretation of these solutions, given that they do not
have the same asymptotic behaviour as proper black holes. This would go beyond the known constructions of microstate geometries for extremal black holes, for
which the solutions only deviate from the classical black hole geometry in a bounded region with the
typical size of the horizon. This is a rather intuitive picture, where one replaces the horizon
by a superposition of microstates, without modifying the dynamic far away from the black hole.
In contrast, if one considers microstate geometries with a different scalar flow, the black hole quantum state will not be well approximated by its classical geometry already at infinity, but the mean value of the scalar flow should nonetheless match the one of the classical black hole solution. We showed in section \ref{sec:BPSbound} that last requirement to be consistent with the asymptotic structure of the solutions of the system, provided the ADM mass satisfies to the extremality bound. In view of the fact that non-extremal black holes
cannot be viewed as isolated systems due to Hawking radiation, the possibility that some fuzziness
at infinity may in fact be desirable, is not ruled out a priori. However, this issue may only
be settled once a sufficiently large class of such hypothetical microstate geometries is constructed.
A first starting point would be to construct solutions to the system defined in this paper, that
include Maxwell--Einstein bases with a spin structure, so that the BPS bound is necessarily satisfied.
In addition, the possibility of more general solutions, which may admit the same asymptotic structure as non-extremal black holes, still remains an open issue.

\section*{Acknowledgment}
We thank Iosif Bena and Nicholas Warner for stimulating discussions.
The work of G. B. was supported by the French ANR contract
05-BLAN-NT09-573739 and the ERC Advanced Grant no. 226371. The
work of S.K. is supported by the European Research Council under
the European Union's Seventh Framework Program (FP/2007-2013)-ERC
Grant Agreement n. 307286 (XD-STRING).

\bibliography{PaperG} \bibliographystyle{JHEP}

\end{document}